# Unmixing Optical Signals from Undersampled Volumetric Measurements by Filtering the Pixel Latent Variables

Catherine Bouchard[1,2], Andréanne Deschênes[1,2], Vincent Boulanger[3,6], Jean-Michel Bellavance[1], Julia Chabbert[1], Alexy Pelletier-Rioux[1], Flavie Lavoie-Cardinal[1,2,4*], Christian Gagné[2,5,7*]

[1]CERVO Brain Research Centre, Université Laval, Québec, QC, Canada

[2]Institute Intelligence and Data, Université Laval, Québec, QC, Canada

[3]Centre for Optics, Photonics and Lasers, Université Laval, Québec, QC, Canada

[4]Dep. of Psychiatry and Neuroscience, Université Laval, Québec, QC, Canada

[5]Dep. of Electrical Eng. and Computer Eng., Université Laval, Québec, QC, Canada

[6]Dep. of Physics, Physics Eng. and Optics, Université Laval, Québec, QC, Canada

[7]Canada-CIFAR AI Chair, MILA-Québec AI Institute

## ABSTRACT

The development of signal unmixing algorithms is essential for leveraging multimodal datasets acquired through a wide array of scientific imaging technologies, including hyperspectral or time-resolved acquisitions. In experimental physics, enhancing the spatio-temporal resolution or expanding the number of detection channels often leads to diminished sampling rate and signal-to-noise ratio, significantly affecting the efficacy of signal unmixing algorithms. We propose *Latent Unmixing*, a new approach which applies bandpass filters to the latent space of a multidimensional convolutional neural network to disentangle overlapping signal components. It enables better isolation and quantification of individual signal contributions, especially in the context of undersampled distributions. Using multidimensional convolution kernels to process all dimensions simultaneously enhances the network's ability to extract information from adjacent pixels, and time or spectral bins. This approach enables more effective separation of components in cases where individual pixels do not provide clear, well-resolved information. We showcase the method's practical use in experimental physics through two test cases that highlight the versatility of our approach: fluorescence lifetime microscopy and mode decomposition in optical fibers. The *latent unmixing* method extracts valuable information from complex signals that cannot be resolved by standard methods. It opens up new possibilities in optics and photonics for multichannel separation at an increased sampling rate.

## INTRODUCTION

Signal unmixing refers to the technique of breaking down a measurement into its unique constituents and their respective proportions[1]. Signal unmixing can apply to any measurements where multiple components are measured at once and need to be separated to be independently resolved. Fluorescence microscopy is one such example, where signals from multiple emitters are often easier to capture together than to separate before detection.

In recent years, major developments have been made to achieve multiplexed fluorescence microscopy, enabling the characterization of cellular and molecular interactions between multiple partners[2–4]. Fluorescence lifetime imaging microscopy (FLIM) uses the arrival time of fluorescence photons to assign them to specific fluorophores or to measure environmental changes in biological samples[5]. The fluorescence decay follows an exponential probability distribution characterized by the parameter $\tau$ corresponding to the mean fluorescence lifetime[6]. Despite careful selection of fluorophores with distinct fluorescence lifetimes, overlapping exponential distributions pose challenges for signal unmixing[7]. The unmixing capacity is limited when multiple fluorophores coexist at the same location by the number of photons acquired at each pixel, by the relative brightness of the fluorescent markers, and by the required image acquisition frequency[4,8–10]. Combining the photons from adjacent pixels can improve the unmixing performance of multiple markers but reduces the effective spatial resolution[4]. For live-cell imaging, where photobleaching and sample movement should be minimized, FLIM measurements are generally associated with low photon-counts (<25 photons per pixel[11]). Similarly, low photon-counts are observed for Stimulated Emission Depletion (STED) super-resolution microscopy combined with FLIM[4,12]. Development of analytical methods that can be applied to limited photon budgets in multiplexed FLIM imaging would have an important impact in the field, allowing to increase sampling rates for live-cell imaging or to resolve nanostructures in super-resolution microscopy.

Different signal unmixing approaches have been proposed specifically for FLIM data analysis. Maximum Likelihood Estimation (MLE)[13] is commonly used for fitting the exponential curve, but it requires high photon counts (>300 photons[14]) for an accurate estimation of the relative abundance of each emitter even for known and well separated lifetime values. Phasor analysis utilizes graphical representations to simplify the identification of contributing lifetimes without requiring iterative fitting[7,15]. Still, emitter separation in FLIM analysis is not accurate for low photon counts (<300) and for more than 3 fluorescent species when using either the phasor or MLE approaches (Methods) [7,16].

Recently, machine learning based solutions have been proposed for FLIM data analysis[17,18]. The *Fluorescence lifetime imaging network* (FLI-Net)[19] is a 3D Convolutional Neural Network (CNN) that takes as input a 3D FLIM volume (x,y,t) where each pixel is characterized by photons sampled in an exponential distribution. FLIMngo[18] uses a well-established object detection architecture (*You Only Look Once*) to predict the average fluorescence lifetime of the fluorophores in each pixel, leveraging both spatial and temporal dimensions. FLI-Net and FLIMngo assign a mean lifetime to each pixel rather than directly unmixing the input image, which is used to study changes in a sample's microenvironment rather than separate multiple components. The FLIM analysis method based on Generative Adversarial Network (GAN) Estimation (flimGANE)[20,21] uses a GAN to generate high-quality FLIM images based on photon-starved (50-200 photons per pixel) measurements. flimGANE is limited to unmixing two components, since the architecture is designed for three outputs: lifetime values $\tau_1$ and $\tau_2$, and relative abundance $A$. Our proposed approach addresses the challenge of unmixing more than two independent fluorophores in FLIM data, tailored for contexts where the photon count per pixel aligns with that observed in STED super-resolution microscopy or timelapse confocal microscopy (<100 photons/pixel).

Microscopy is not the only experimental context where measurements are noisy or undersampled. All physical measurements are expected to be contaminated with a certain amount of noise that is either generated from the instrument, from external sources, or from random statistical fluctuations. Measurements are generally limited in frequency, which can also lead to noisy or under-defined signals. Accurate component separation is challenging when the specific signal distribution of each component is unknown, or when the sampling coverage is insufficient[22,23]. For multidimensional measurements, such as volumetric or spatio-temporal data, the interdependence between different dimensions can be effectively leveraged to create a richer representation[24] that facilitates unmixing. An efficient unmixing method should capitalize on the connectivity between signals measured at different points, such as neighboring pixels in images. Data-driven approaches excel in feature extraction over multiple dimensions[24]. A data-driven unmixing algorithm can learn to recognize the spatial features that are indicative of distinct structures and deduct which parts of different signals should be combined together for an information-rich representation of undersampled pixels. To verify whether our unmixing method can be applied to other types of measurements and temporal distributions, we developed a use-case for a type of distribution that differs from the exponential decays of FLIM: the decomposition of overlapping propagating modes from the intensity signal at the output of a multi-mode optical fiber. This simulated experimental context allows for the evaluation of the generalizability of our unmixing method.

Multi-mode optical fibers (MMF) support the simultaneous propagation of multiple light modes. Spatially and spectrally resolved ($S^2$) imaging[25] is a measurement commonly used for characterizing how light travels through multi-mode optical fibers. By measuring the spatial distribution of the intensity at the output of the fiber for a range of laser wavelengths, the propagation modes of the light through the fiber can be characterized. The measurement at the fiber output is not a simple linear combination of each distinct mode due to interference. When multiple modes propagate through the fiber, the intensity oscillates as a composite of sinusoidal functions, each corresponding to the interaction between pairs of modes[25].

In an $S^2$ measurement for which the sampling frequency is above the Nyquist criterion, the measured oscillations allow to separate each mode with a Fourier transform based method[26] or by principal component analysis (PCA)[27,28]. Other methods exist to achieve real-time mode decomposition from a single image[29–31], but they all require the *a priori* knowledge of the spatial distribution of each mode. To our knowledge, no method for mode decomposition in multi-mode fibers allows to separate modes with unknown spatial distributions and successfully retrieve their contribution in measurements using a sub-Nyquist sampling rate.

In this work, we propose a computer vision inspired unmixing method that applies to various types of undersampled multidimensional optical measurements, outperforming signal analysis curve-fitting based methods. Our method, called *Latent Unmixing*, uses learned spatial features to transform the overlapping input contributions to a latent space where they are untangled and can be separated by applying linear bandpass filters directly in the latent space (Fig. 1). It offers a versatile approach to the unmixing problem that can be easily adapted to different distributions and number of unmixed targets. We demonstrate the applicability of our method using two use cases in experimental physics: FLIM and mode decomposition in multi-mode optical fibers.

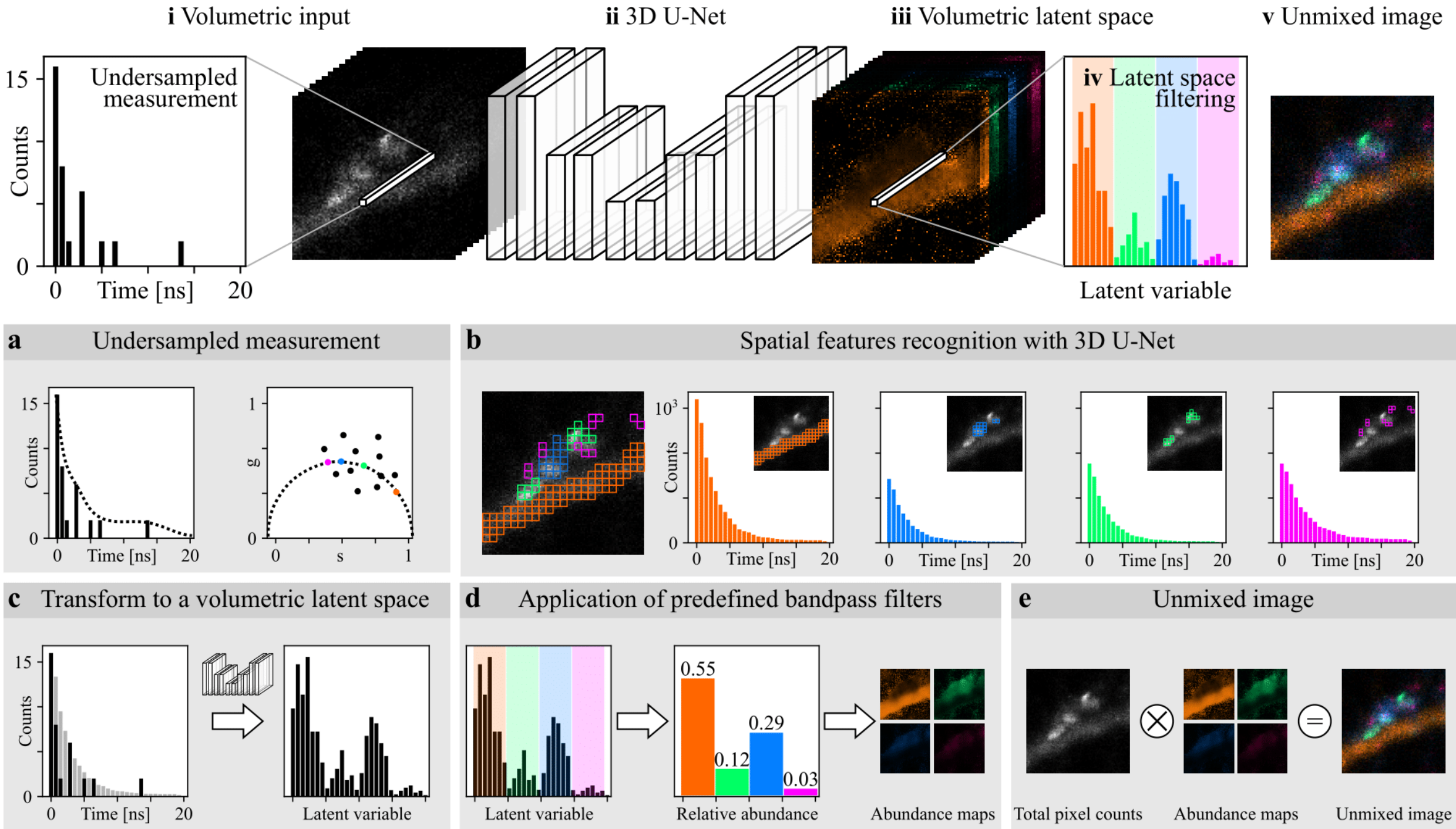

**Fig. 1 - The *Latent Unmixing* method.** The *Latent Unmixing* method maps volumetric undersampled measurements (**i**) using a 3D U-Net (**ii**) to a latent space (**iii**) where pre-defined bandpass filters (**iv**) are applied to retrieve the relative abundance of each contributing component (**v**). **a,** Methods relying on well-defined distributions (left; curve-fitting, right; phasor analysis) cannot retrieve the relative abundance of multiple components from undersampled distributions. **b,** The 3D U-Net is trained to recognize spatial features indicative of distinctive structures. The resulting representation combines the information from the pixel ensemble corresponding to continuous structures in the image, instead of the pixel-wise undersampled input. **c,** The information decoded from the spatial features is exploited to project the distribution of each pixel to the latent space. **d,** Pre-defined filters are applied to the latent representation to extract the abundance of each component. **e,** The measured counts are multiplied with the abundance maps, resulting in the predicted unmixed image.

## RESULTS

In CNNs, the latent space simplifies the raw input pixel data into an informative representation that is used by the subsequent layers to perform the task[32] (Fig. 1**iii**). Our *Latent Unmixing* method uses this abstract representation of the data to translate the 3D input into a multi-channel 2D output. *C* predefined bandpass filters are applied to the pixel latent space of the 3D U-Net output to pool this pixel latent dimension into *C* relative abundance maps (Fig. 1**iv**). This forces the network to learn a mapping that transforms the input overlapping distributions into a latent space where they do not overlap and can be separated using the predefined bandpass filters. The *C* relative abundance maps are multiplied to the amplitude of the initial measurement (photon counts or intensity) to obtain a *C*-channel multiplexed image (Fig. 1**v**).

We first demonstrate the performance of the *Latent Unmixing* method on a variation of the MNIST dataset, then on two simulated experimental physics use cases: FLIM and mode decomposition in optical fibers. In all three cases, the input data consists of noisy, undersampled, and overlapping signal distributions to be decomposed into multi-source outputs.

**3D MNIST with simulated time dimension**

We created the 3D-MNIST[33] dataset to benchmark the performance of the *Latent Unmixing* method. This dataset consists of volumes where images of the digits 4, 8 and 9 are overlapped in the spatial dimension, with a third dimension simulating a decay probability distribution for each class of digit (Methods, Algorithm 2). The *Latent Unmixing* method was trained over the 3D-MNIST dataset to unmix overlapping digits by mapping the input distributions (Supplementary Fig. 3, top-left) to a pixel latent space where the contributions are well separated to fit predefined bandpass filters (Supplementary Fig. 3, Fig. 2**c**). Four bandpass filters each of width 7, equally spaced along the pixel latent vector of length $L$=28 were used for this dataset (Methods).

The 3D-MNIST dataset is designed to have a low signal-to-noise ratio, such that the input signal distributions do not provide enough counts to enable the application of curve fitting methods for extracting the relative abundance of up to four components (Supplementary Fig. 5). The *Latent Unmixing* model projects the undersampled distributions to the latent space in a way that untangles their relative abundance (Supplementary Fig. 3, Fig. 2**c**, Methods). Because it is built from a 3D-UNet, the *Latent Unmixing* method combines the spatial features indicative of each digit along with the pixel-wise signals to accurately assign the different contributions. The spatial features indicative of each digit (4, 8, 9) are correctly replicated in the predictions, leading to an accurate pixel-wise reconstruction of each channel (Fig. 2**e**). The *Latent Unmixing* method achieves a Pearson correlation coefficient between true and predicted pixel counts above 0.69 for all channels over the test set (Fig. 2**f**). To confirm that the third dimension is used by the network to predict the unmixed image, we compared the result with a 2D U-Net trained on the same set of images of overlapping digits without the temporal dimension. The digits cannot be recognized in the predicted unmixed images, and the predicted and true counts are not correlated (Methods, Supplementary Fig. 4).

The performance of the *Latent Unmixing* method is compared with the established curve fitting approach MLE[13] to unmix the same test set pixel-by-pixel (Methods). In this context with limited counts per pixel, MLE cannot separate the different contributions from the 4 channels (Table 1, Methods)[13,14,34]. In our experimental settings, the average over all pixels of the test set is 5.6 counts per pixel, which is a full order of magnitude below the described limits of MLE[13,14].

| | **MLE** | ***Latent Unmixing*** |
|---|---|---|
| Channel 4 | 1.98 | 0.324 |
| Channel 8 | 4.76 | 0.647 |
| Channel 9 | 4.88 | 0.553 |
| Background | 11.69 | 1.01 |

**Table 1 -** Mean squared error (MSE) between true and predicted counts using MLE and *Latent Unmixing* over the test set of the 3D-MNIST dataset (N=800 images).

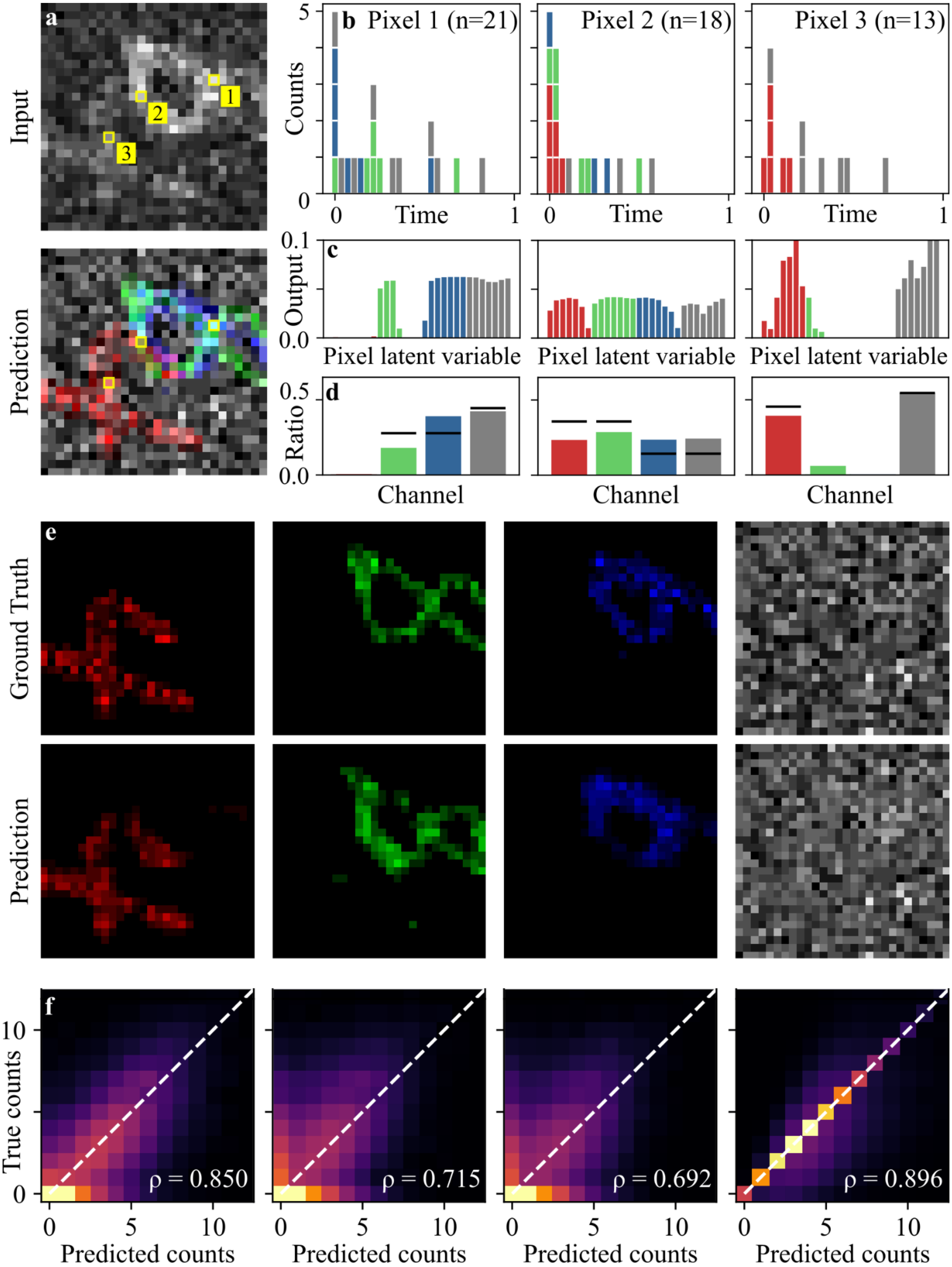

**Fig. 2 - *Latent unmixing* on the MNIST Dataset. a,** Input intensity image (top) and the corresponding unmixed 4 channel prediction (bottom). **b,** The input distributions for the *n* photons in each of the three pixels highlighted in **a** (yellow boxes). The photons from the digits 4, 8, 9 and the background are respectively represented by the colors red, green, blue and gray. **c,** Pixel latent variable (output of the 3D U-Net) for the same three pixels. See Supplementary Fig. 3 for the pixel latent variable averaged over the whole test set for each channel. **d,** Ratios obtained after applying bandpass filters on the pixel latent variable. The black horizontal lines represent the relative abundance of each channel in the ground truth image. **e,** Channel-wise unmixing of an example image from the test set, with ground truth (top) and *Latent Unmixing* prediction (bottom). **f,** Correlation matrices between true and predicted photon counts (bottom) (N=800 images, 627,200 pixels). Dashed white lines show a perfect correlation with a slope of 1. Colormaps are square-root normalized and scaled to the second maximum of each matrix. ρ is the Pearson correlation coefficient between the ground truth and predicted counts for each channel.

### Fluorescence lifetime microscopy

The STED-FLIM unmixing task serves as a demonstration of the *Latent Unmixing* method in a real experimental setting. For multi-channel STED-FLIM, channel separation is challenging at low photon counts (<50 photons per pixel), for more than two fluorescent markers, and for small lifetime differences between the emitters (<1 ns)[8,9,11]. To ensure the generalizability of the model over different spatial features, we trained a conditional U-Net with a conditioning vector inspired by the denoising diffusion model[35] (Fig. 3**b**). For conditional unmixing, the U-Net takes as input a vector $\mathcal{T} = [\tau_1, \tau_2, \tau_3, \tau_4]$ in addition to the volumetric image to unmix. For images with less than 4 channels, the empty channels are represented by $\tau$=-1 (Methods). The *Latent Unmixing* model was trained on the STED-FLIM dataset to map the input volumes (depth $L$=40 time-bins, width 0.5 ns) (Fig. 3**a**) to a latent space of the same length, on which 4 bandpass filters equally spaced and of equal width of 8 were applied (Fig. 3**c-e**), similarly to the 3D-MNIST experiment. The STED-FLIM dataset is generated from real 4-channel fluorescence microscopy images from the *4-channel STED Dataset*[36] for which a simulated lifetime is attributed to each individual photon (Methods). Unlike the 3D-MNIST dataset, the STED-FLIM dataset does not include an independent background channel with a specific lifetime. To create the background channel, we assign all pixels with a count below 4 photons to the background. Because background photons have an unknown lifetime distribution, we assign the 6 bins at the intersection between channel-filters and the 2 bins at the extremities of the latent vector to a fifth channel corresponding to the background photons (Fig. 3**c-e**, Methods).

The 12 test images from the *4-channel STED Dataset*[36] were each split into 256 × 256 crops using a sliding window with steps of 128 pixels resulting in 800 testing crops. Each crop was processed by the *Latent Unmixing* model independently. The predicted crops were recombined to generate the full field of view image by using the central 128 × 128 region of each prediction without overlap, thereby eliminating potential border effects.

Generalizability to 2-, 3-, and 4-channel unmixing

The vector $\boldsymbol{\mathcal{T}}$ encoded into the U-Net makes the network adaptable to different numbers of emitters between 2 and 4. The projection of the input volume in the latent space adapts to the number of emitters dictated by $\boldsymbol{\mathcal{T}}$ (Fig. 3**a-b**). When the 4 values of $\boldsymbol{\mathcal{T}}$ are positive, the network decodes the input into 4 channels (Fig. 3**c**). Setting the last value of $\boldsymbol{\mathcal{T}}$ to -1 excludes signal from the last bandpass filter, yielding a three-channel image (Fig. 3**d**). Fixing the last two values of $\boldsymbol{\mathcal{T}}$ at -1 similarly limits the signal to two channels (Fig. 3**e**). An input volume can be unmixed into 2-, 3-, or 4-channel using a single 3D U-Net trained once, simply by changing the values of $\boldsymbol{\mathcal{T}}$ given as input along with the FLIM volume (Fig. 3).

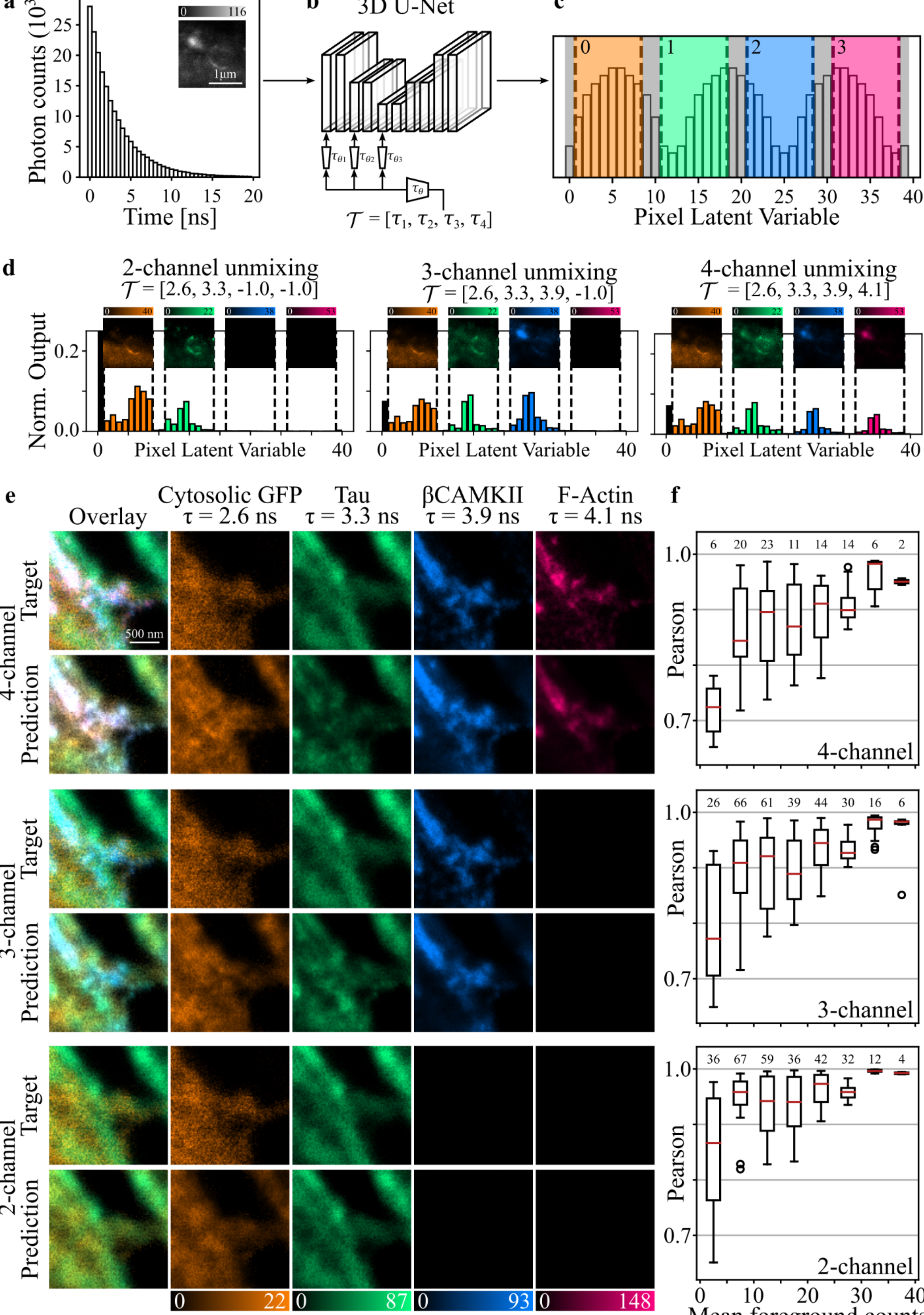

**Fig. 3 - Unmixing simulated fluorescence lifetime microscopy images. a,** Photon arrival time distribution for all pixels in an example input crop (top right). **b,** The 3D U-Net used for conditional unmixing also takes as input the vector $\boldsymbol{\mathcal{T}}$ that contains the lifetime values of the 4 emitters in the sample, in ascending order. If the sample contains less than 4 emitters, the lifetime values of the empty channels are replaced by -1.0 and ordered last. $\tau$ is encoded by a multilayer encoder $\tau_\theta$ and layer specific encoders $\tau_{\theta 1}$, $\tau_{\theta 2}$, and $\tau_{\theta 3}$ (Methods). **c,** Predefined bandpass filters, with four channels represented in colors and the discarded background photons in gray. **d,** Output for 4-color unmixing with $\boldsymbol{\mathcal{T}}$ = [2.6, 3.3, 3.9, 4.1], for 3-color unmixing with $\boldsymbol{\mathcal{T}}$ = [2.6, 3.3, 3.9, -1.0], and for 2-color unmixing with $\boldsymbol{\mathcal{T}}$ = [2.6, 3.3, -1.0, -1.0]. Lifetime values are chosen to reflect the theoretical lifetimes of the fluorescent dyes used in this specific experimental context (ATTO 490LS, STAR635P, Alexa Fluor 594, Alexa Fluor 488). The displayed colorscale of each individual image is the 99.5th percentile of the 4-color prediction. Images show 4 fluorescently labeled neuronal proteins. Orange: Microtubule-associated protein 2 (MAP2), Green: Postsynaptic density protein 95 (PSD-95), Blue: Vesicular glutamate transporter 1 (VGLUT1), Pink: RNA-binding protein fused in sarcoma (FUS). **e,** Example crops from the test set with simulated lifetime for a set of four fluorescently labeled proteins (cytosolic Green Fluorescent Protein (GFP), Tubulin Associated Unit (Tau), β-$Ca^{2+}$/calmodulin-dependent protein kinase II (βCaMKII) and F-Actin in fixed cultured hippocampal neuronal cultures). The ground truth has either 4 (top), 3 (middle) or 2 (bottom) channels to unmix. Color scales at the top show min-max display values and correspond to the 99.5th percentile of the ground truth channel. **f,** Pearson correlation coefficient between the pixel counts in the ground truth and the *Latent Unmixing* prediction for 2- (left), 3- (middle) and 4-channel (right) unmixing. Images are binned into ranges of 5 counts per pixels over the foreground pixels of the image. The numbers above the boxes are the number of channels in each distribution. N=288 for 2 channels, N=288 for 3 channels, and N=96 for 4 channels (Methods).

The generalizability of the *Latent Unmixing* method over 2-, 3-, and 4-channel unmixing was evaluated over a test set of 264 FLIM images with 2, 3, or 4 simulated emitters produced from 12 4-colors intensity images (Methods). For interpreting the performance over the foreground pixels, a foreground mask was computed using a triangle thresholding approach applied independently to each channel. For very low photon counts (less than 5 photons per foreground pixel), the *Latent Unmixing* method already achieves a median Pearson correlation coefficient of 0.85 ± 0.10 (mean

± standard deviation) for 2-channel, 0.79 ± 0.10 for 3-channel, and 0.72 ± 0.05 for 4-channel unmixing (Fig. 3**f**). For slightly higher counts (5 to 10 photons per pixel), the performance increases abruptly, and is then relatively stable for up to the maximum mean count (35 to 40 photons per pixel) for 2-, 3- and 4-channel unmixing (Fig. 3**f**). Because it interprets the spatial features of the image as well as the time-dimension, the *Latent Unmixing* method requires lower photon counts to make accurate predictions, as opposed to curve fitting based methods where the accuracy correlates with the number of photons.

The *Latent Unmixing* method outperforms the phasor analysis method over the test set for 2-, 3-, and 4-channel unmixing for all combinations of lifetime values tested. For photon counts that are realistic of what is expected from experimental STED-FLIM acquisitions (counts in the range 0 to 50 photons per pixel), the projection of the pixels to the phasor plot space is too noisy for the abundance to be accurately predicted, resulting in noisy and unreliable unmixed channels (Fig. 4 and Supplementary Fig. 10). With the *Latent Unmixing* method, the spatial features derived from the neighborhood of each pixel helps reach more accurate predictions, achieving a Pearson correlation coefficient (mean ± STD) of 0.88 ± 0.08 over the 4 channels of the 12 test images with simulated lifetime values of 1.0, 2.0, 3.0 and 4.0 ns, compared to 0.52 ± 0.03 using the phasor analysis method.

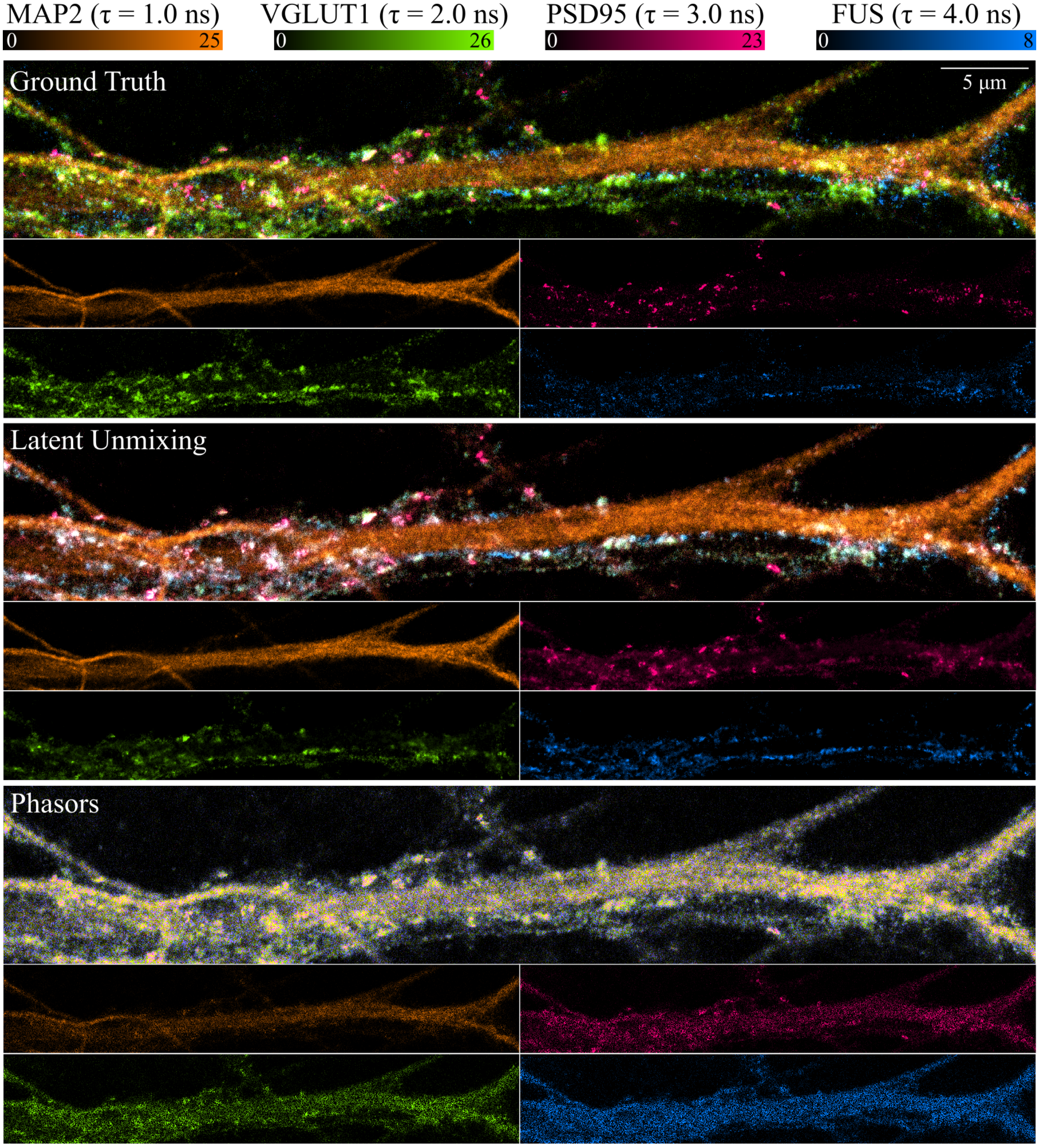

**Fig. 4 - 4-channel unmixing with the *Latent Unmixing* and the phasor analysis methods**. Photon counts are representative of real acquisitions, as they are taken directly from the acquired intensity confocal and STED images. All colormaps are scaled from 0 to the 99.5th percentile of the ground truth channels, indicated by the colorbars. Lifetime values (1.0 ns, 2.0 ns, 3.0 ns and 4.0 ns) are chosen to serve as an easy case ($\Delta\tau$ = 1.0 ns for all channels), while remaining in the realm of possible lifetime values with current commercially available dyes. Refer to Supplementary Fig. 8 and Supplementary Fig. 9 for more examples over different protein combinations.

Robustness to low-photon counts

A key aspect of the *Latent Unmixing* method is that the information from neighbouring pixels is processed simultaneously by the 3D convolutional kernels, which should lead to more accurate unmixing results for low-photon counts than single-pixel methods like the phasor analysis. Since the phasor analysis method cannot accurately unmix more than two emitters in the photon counts range we operate at (Supplementary Figs. 8-10), the quantitative comparison between the two methods concentrates on the 2-channel case. The quantitative analysis covers a test set of 168 2-channel FLIM images simulated from 12 real intensity STED images and 14 combinations of two lifetime values (Methods). For all $\Delta\tau$ tested, the relative error between the true and predicted counts per channel for each pixel decreases when the photon count increases (Fig. 5**d**). For lifetime differences below 2.2 ns, *Latent Unmixing* outperforms the phasor analysis method for both the image-wise Pearson correlation coefficient (Fig. 5**c**) and the pixel-wise relative error (Fig. 5**d**). The effect is more pronounced for lower photon counts. For all lifetime differences covered (0.2 - 2.8 ns), the performance of *Latent Unmixing* is more stable than the phasor approach for counts below 50 photons/pixel.

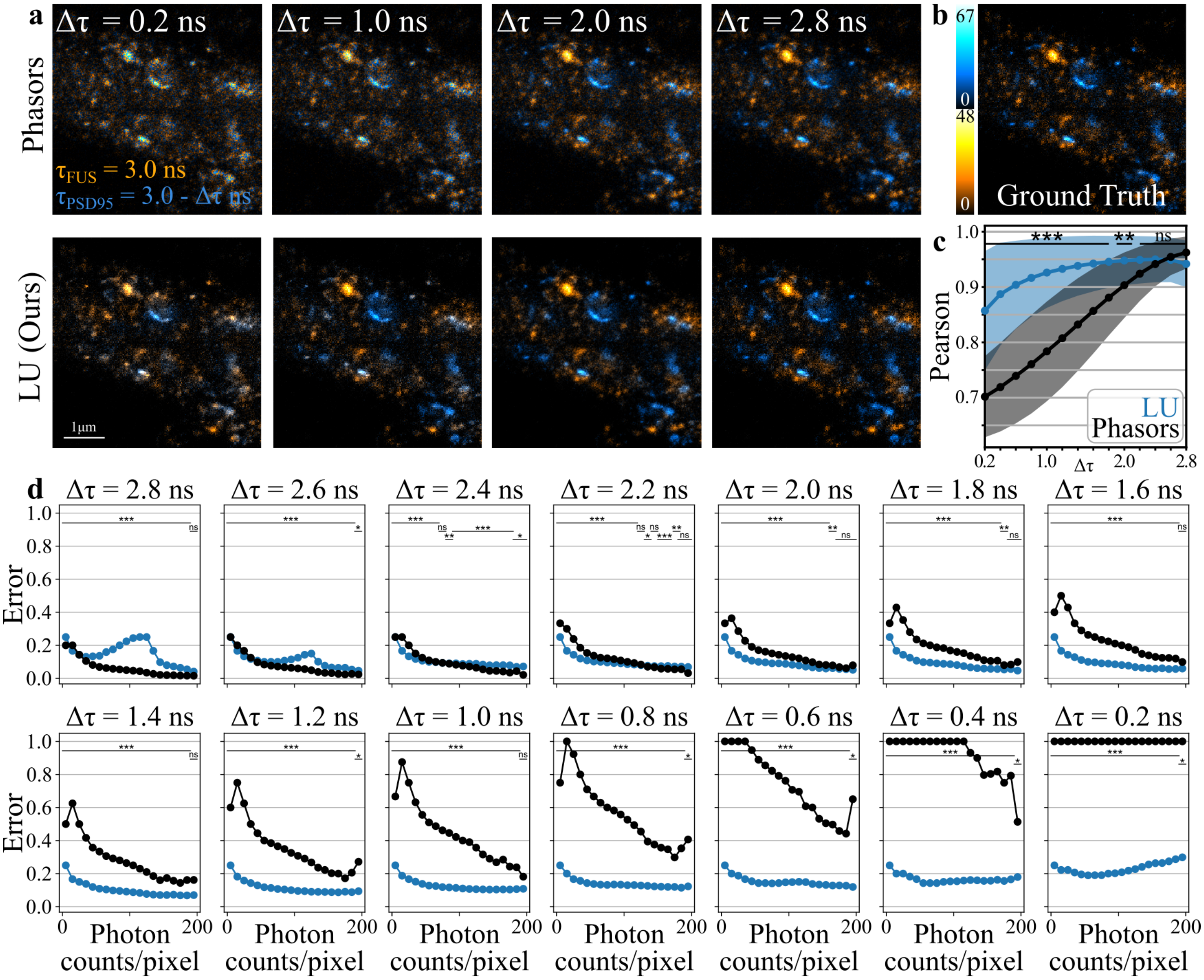

**Fig. 5 - a,** Image crop from the STED-FLIM test dataset comparing the performance of the phasor analysis and our *Latent Unmixing* (LU) method as a function of the difference between the lifetime values of both channels ($\Delta\tau$). **b,** Ground truth. The colorbars show the display range for all images in **a** and **b**. **c,** Pearson correlation coefficient computed over the images of the test set using the *Latent Unmixing* and the phasor analysis methods. Each point is the mean computed over the 2 channels of the 12 test images (N=24), and the shaded region is the standard deviation. See Supplementary Fig. 12 for an additional example. **d,** Influence of the lifetime difference and the photon counts for 2-channel unmixing. Median relative error for the predicted counts of all pixels above 3 photons/pixel for the test images, using the predictions of the phasor analysis (black) and the *Latent Unmixing* (blue) methods. Pixel counts are binned in 10-count intervals. From left to right and top to bottom, the difference between the two lifetime values decreases from 2.8 ns to 0.2

ns. For large differences, both methods perform similarly. For smaller differences (≤ 1.6 ns), *Latent Unmixing* outperforms the phasor method, with the difference between the two methods increasing for decreasing photon counts. Statistical analysis: **c,** the two distributions have the same mean, **d,** the two distributions come from populations with the same median (Methods), Statistical analysis null hypothesis: the two distributions have the same mean (Methods), *** : p≤0.001, ** : p≤0.01, not significant (ns): p>0.05.

Robustness to close lifetime values

Finding pairs of fluorophores that have distant lifetime values can be a challenge for biological applications. The ability to correctly unmix two channels with lifetime values that are close apart (<1 ns) is therefore a valuable advantage for an unmixing method. Our *Latent Unmixing* method shows better robustness to close lifetime values than the phasor analysis method (Fig. 5**a-c**), achieving a Pearson correlation of ρ=0.86 ± 0.11 (mean ± STD, N=24) for lifetime values only 0.2 ns apart, compared to the phasor analysis method that achieves only ρ=0.68 ± 0.06. In the 2-channel unmixing context, differences of 2.4 ns and above lead to similar performances for both approaches. In practice, a lifetime difference above 2.2 ns is rarely achievable given the finite number of fluorescent dyes available commercially.

**Unmixing real FLIM measurements**

We applied the same *Latent Unmixing* model trained on the simulated STED-FLIM dataset to real FLIM measurements (Fig. 6) with different lifetime values, photon counts, and biological structures. The real measurements, sampled in the *Confocal- and STED-FLIM images of neuronal proteins dataset*[37], require only simple preprocessing steps to fit the format of the simulated training data. The instrument response function is first removed by shifting the exponential distribution such that its maximum is at the 0-value, commonly known as tail-fitting[38]. The time measurement is then binned into 40 intervals of 0.5 ns to match the training data.

Control samples with only one fluorescently tagged protein were imaged to measure the average lifetime of each individual fluorophore. The lifetime value of each fluorophore was found using MLE-based curve-fitting over the time-dimension of the sum of all pixels in the control image (Methods). The sum of all collected photons over all pixels of a control image is between $6.7 \times 10^4$ and $3.4 \times 10^6$ counts, ensuring MLE is a suitable method for lifetime estimation. For each sample, 12 to 20 control images were used and the measured fluorescence lifetime was averaged over all control images (Supplementary Table 9).

To validate the predictions, we sum two control images along the spatial dimension to generate a two-channel ground truth image (Fig. 6**a**). *Latent Unmixing* can unmix real FLIM measurements, even though it was trained only with a simulated time dimension sampled from simple exponential decay distributions. The lifetime conditioning also ensures the method is adaptable across a range of lifetime values and can therefore be applied to various combinations of emitters without requiring any retraining or fine-tuning (Fig. 6**b-c**).

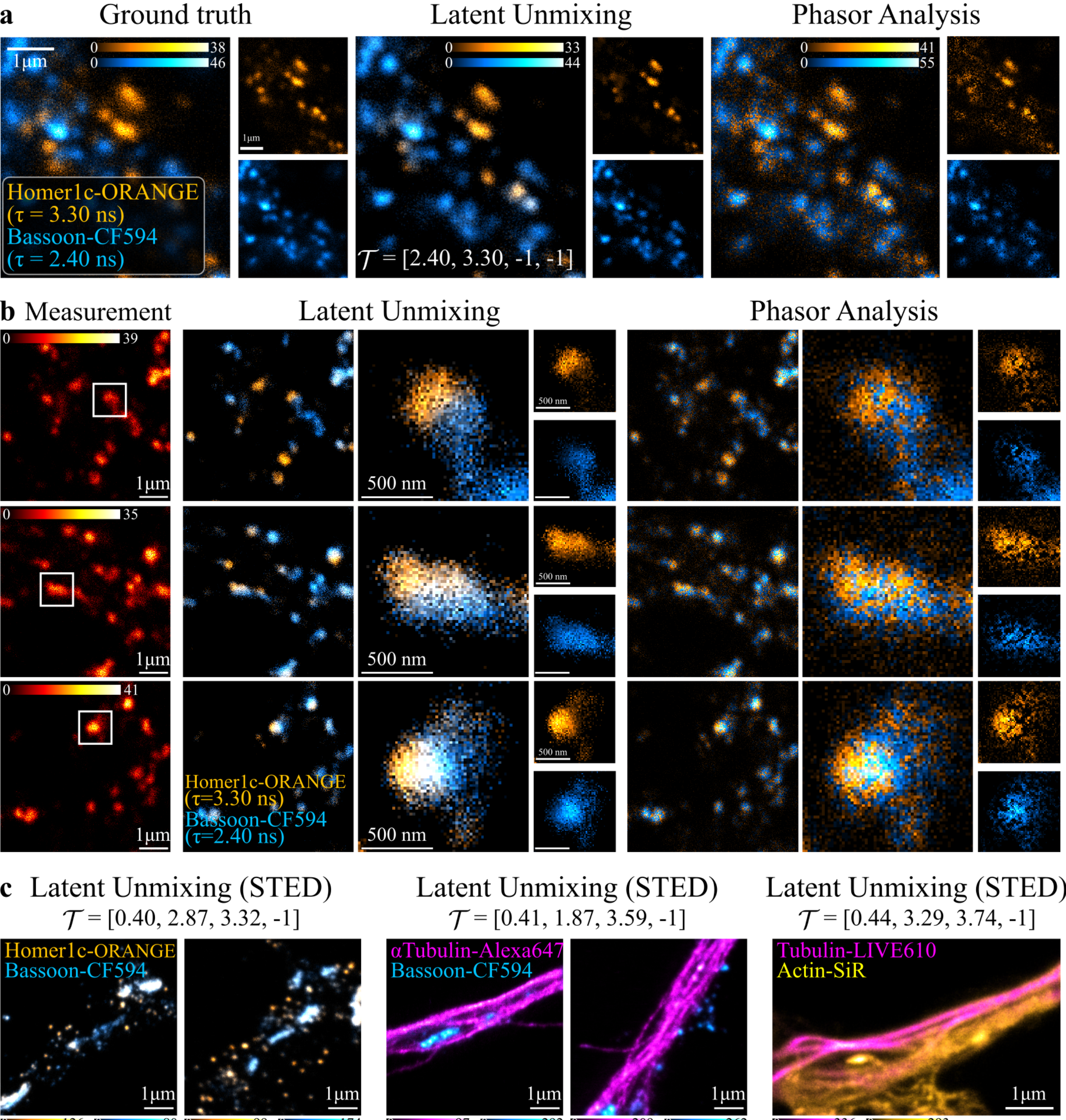

**Fig. 6 - Unmixing real FLIM measurements. a,** Unmixing of FLIM measurements with phasor analysis and *Latent Unmixing* of Homer1c tagged with STAR ORANGE ($\tau$=3.74 ns) and Bassoon with CF594 ($\tau$=2.67 ns), $\Delta\tau$=0.90 ns, both imaged with the confocal-FLIM modality. The 2-channel mixed images are generated by adding together two 1-channel control images, so that the results can be compared with a perfect ground truth. **b,** Results on real FLIM measurements with phasor analysis and *Latent Unmixing* of the same synaptic protein pair imaged with the confocal-FLIM

modality. Images on the right are zoomed in regions identified by the white boxes in the mixed intensity images. **c,** *Latent Unmixing* results on real FLIM measurements for three pairs of fluorescent dyes imaged with the STED-FLIM modality. From left to right: Homer1c tagged with STAR ORANGE ($\tau$=2.87 ns) and Bassoon with CF594 ($\tau$=3.32 ns), $\Delta\tau$=0.45 ns; $\alpha$Tubulin tagged with Alexa Fluor 647 ($\tau$=1.87 ns) and Bassoon with STAR635P ($\tau$=3.59 ns), $\Delta\tau$=1.72 ns; Tubulin tagged with LIVE610 ($\tau$=3.74 ns) and F-Actin tagged with SiR-Actin ($\tau$=3.39 ns), $\Delta\tau$=0.45 ns. The first value of the vector $\boldsymbol{\mathcal{T}}$ is used to predict an additional channel (not shown) corresponding to the shorter lifetime photons emitted in the periphery of the STED beam (Supplementary Table 10, Methods)[39]. All colormaps are scaled to the maximum intensity of the corresponding channel.

**Mode decomposition in multi-mode optical fibers**

We next applied the *Latent Unmixing* to the unmixing of interfering sinusoidal signals propagating in a multimode fiber. The Spatially and Spectrally resolved Multi-Mode optical Fibers ($S^2$-MMF) dataset consists of simulated images of the intensity measured at the output of optical fibers for different input wavelengths[25]. The simulated measurements have two spatial dimensions (2D images of the light distribution at the fiber output, Fig. 7**a**) and one spectral dimension (Fig. 7**b**). The dataset was computed with previously described equations[27].

The $S^2$-MMF dataset is used to test if *Latent Unmixing* can generalize to other types of distributions than exponential decays; here, interfering sinusoidal signals with varying spatial intensities. For every single pixel, the sampling rate is below the Nyquist frequency, which is the minimal frequency of acquisition that would allow the application of Fourier-based processing methods. By applying our approach to the $S^2$-MMF volumes, the correlation from multiple pixels provides information to find the unknown intensity and spatial distribution of multiple coherent electromagnetic modes propagating simultaneously in a given optical fiber. The *Latent Unmixing* method was trained to map the input data to a latent space of length 60, where 6 bandpass filters of width 10 are applied to separate the space into 6 target modes (Fig. 7**c**). Only the volumetric simulated measurement is given as input to the network; the model is not conditioned over the characteristics of the simulated optical fiber. The ground truth is the relative spatial intensity of each mode at the central wavelength (1550 nm), and the channels are ordered in ascending mode order.

The results obtained with the S²-MMF experiment demonstrate that the *Latent Unmixing* method can untangle the different signal contributions from sub-Nyquist measurements (Fig. 7). We first evaluate the relative abundance maps of each mode. *Latent Unmixing* correctly recreates the spatial features of each mode, which varies in function of the geometric parameters of the simulated fiber (Fig. 7**d**, Supplementary Fig. 15). The correlation between the pixel-wise values of the ground truth and predicted pixels is above 0.778 for all modes (Supplementary Fig. 14, Supplementary Table 11). During training and at inference time, the model is not conditioned over the geometric characteristics of the simulated optical fiber; the spatial distribution of the modes is predicted directly from the input signals. This result makes the *Latent Unmixing* method more generally applicable than methods that require the spatial distributions to be known beforehand to correctly retrieve the relative intensity of each mode[29–31].

The fraction of the total intensity that is propagated through each mode is one of the aspects analyzed when using S² measurements to characterize optical fibers[25]. This relative intensity is obtained by summing the pixel values of the predicted image of each mode, and dividing by the pixel values over all 6 predicted modes. The predicted and ground truth fractions of intensity correlate (Supplementary Table 12, Fig. 7) over all numbers of propagating modes. There are more errors (outliers from the diagonal in Fig. 7) for smaller numbers of modes in the mixture (2 or 3 modes) than for higher numbers of modes (4, 5 or 6), because a small relative intensity is attributed to non-propagating modes. This is a drawback from the fact that our method does not require knowledge about the spatial distribution of the modes, the number of modes in the mixtures, or the geometric parameters of the optical fiber. For an experimental context where those parameters are completely or partially known, a conditioning branch (Supplementary Fig. 1) could be added to improve the performance, as it was done for the STED-FLIM unmixing experiments.

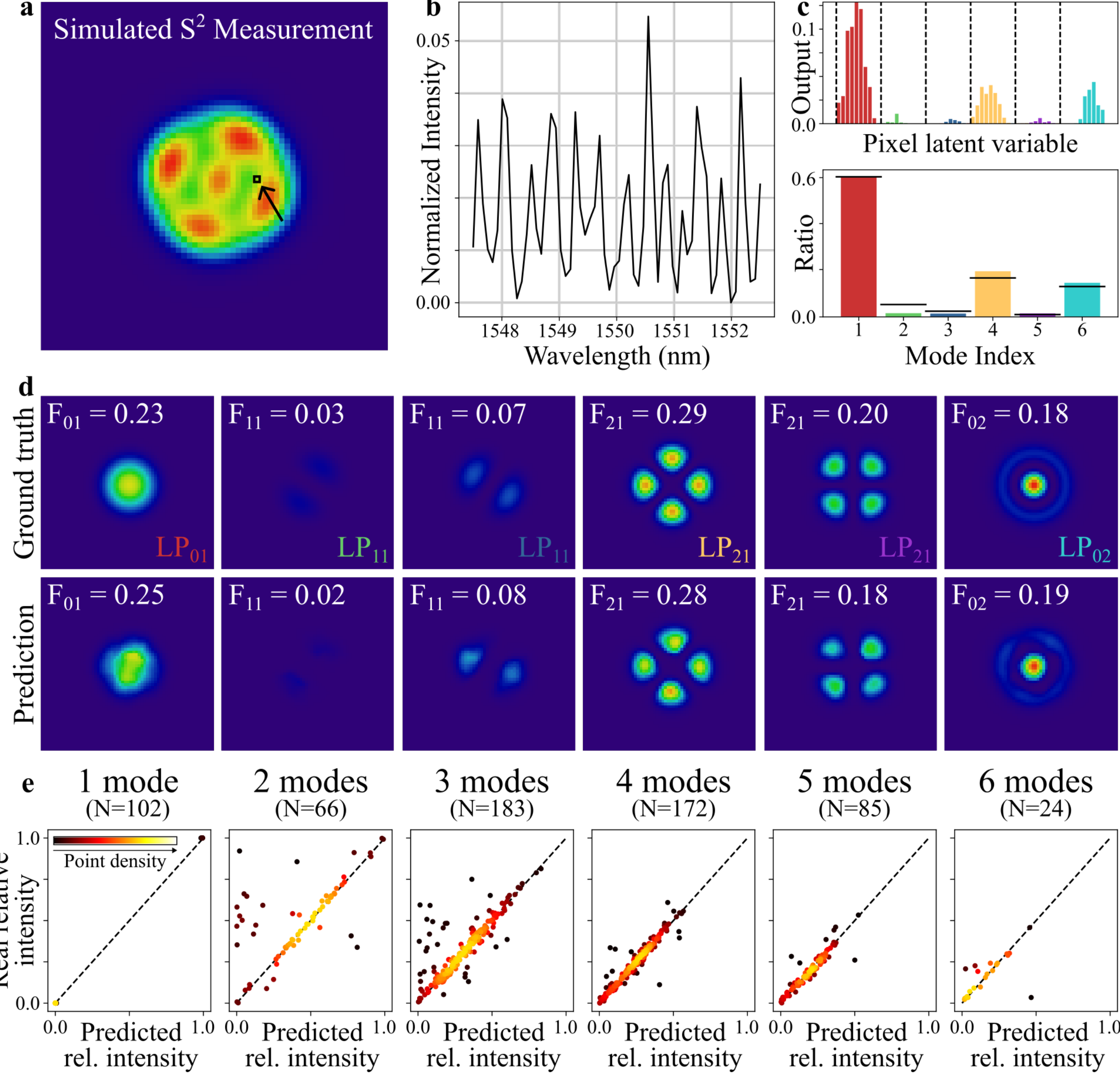

**Fig. 7 - Mode decomposition with *Latent Unmixing*. a,** Input spatial distribution of a test $S^2$ measurement. **b,** Spectral distribution of the highlighted pixel. **c,** Top: raw distribution of the same pixel in the latent space. Bottom: distribution pooled into 6 classes for the 6 modes, with black lines indicating the ground truth proportion of intensity in each mode. **d,** Ground truth (top) and prediction (bottom) for the 6-mode decomposition of the test data shown in **a** and **b**. More examples of predicted spatial distributions are shown in Supplementary Fig. 15. **e,** Correlation between the target and predicted intensity over different mixtures of propagating modes. Dashed line is the perfect correlation with a slope of 1. Points are color-coded by the spatial density of points,

estimated by kernel-density with Gaussian kernels, to show the higher density over the diagonal. For 2 and more modes, points for which the ground truth channel is empty are omitted for clarity (see Supplementary Fig. 13 which includes all points).

## DISCUSSION

The *Latent Unmixing* method is a deep learning technique for decomposing mixed images into their individual contributing components, drawing on computer vision to address signal-processing challenges in experimental physics. It relies on a 3D CNN to capture and combine features from different dimensions simultaneously, leading to improved representations of complex patterns in the three-dimensional space[40].

*Latent Unmixing* uses a 3D U-Net to combine the spatial and time/spectral dimensions in contexts where, if taken individually, each dimension (spatial and time/spectral) does not include enough information to allow correct separation of the components. By integrating these spatial relationships, more information is extracted, achieving more accurate unmixing results than conventional methods that rely only on the time/spectral dimension. Our approach takes advantage of an already demonstrated neural network architecture, but the architecture choice is not a limiting factor of the method: it could easily be adapted to other architectures.

Our approach is first illustrated and validated on a toy synthetic dataset based on MNIST, where digit images are juxtaposed, generating the time dimension with different configurations for each of the digits used. This toy dataset confirmed the validity of the *Latent Unmixing* method by demonstrating its superiority over a pixel-wise curve fitting method (MLE, Methods) and a spatial-only method (2D U-Net, Methods). We then demonstrated the *Latent Unmixing* method for two experimental contexts: FLIM and mode decomposition in optical fibers, exemplifying the generalizability of the method to different signal distributions and experimental settings. In both cases, our method adapts to class imbalance across data samples (number of photons per pixel in FLIM and number of propagating modes in MMF). In this study, we crafted datasets that accurately represent real experiments, while providing an associated ground truth to evaluate the performance of the proposed *Latent Unmixing* approach. Our dataset format can easily be adapted to fit specific parameters, distributions, number of channels, and ranges of lifetime or wavelength values. Even though the *Latent Unmixing* method was trained with a time dimension completely simulated from sampling in exponential decay functions, the results on real FLIM measurements support that the method can be applied to real lifetime distributions associated with FLIM experiments.

In conclusion, the *Latent Unmixing* method takes advantage of the capacity of deep multidimensional convolutional neural networks to combine the information from all dimensions, which allows unmixing undersampled and noisy data. This work was concentrated on 3D measurements, but future work could explore the potential of multidimensional CNNs for *Latent Unmixing* over higher numbers of dimensions, e.g. spectral-FLIM microscopy[41] ($x,y,\lambda,\tau$). Our method was crafted in a way that is generalizable to different numbers of target channels (1-4 for 3D-MNIST, 2-4 for STED-FLIM, 1-6 for $S^2$-MMF) and different types of distributions (exponential decay curves, interfering sinusoidal signals). This novel use of deep neural networks showcases the capacity of deep networks to process sparse information in its entirety, which is of great interest for a range of imaging applications.

## METHODS

### The 3D U-Net Model

The U-Net model[42] and its 3D version[43] are CNNs that were initially designed for the segmentation of biomedical images. We exploit a 3D U-Net[43] model to map the optical data from the input space, where the signal components are undersampled and overlapping (Fig. 1**i**), to a latent space where the different contributions are differentiated. The 3D kernels process the three dimensions of an input volume simultaneously, thus combining essential information from neighbouring pixels and neighbouring time- or spectral-bins (Fig. 1**ii**). The network has 4 downsampling blocks and 4 upsampling blocks (Supplementary Fig. 1). To generalize the model to contexts where the input distribution parameters vary, we developed a conditional paradigm where the parameters that define the distribution of the input volume are embedded into the encoder blocks. The details of the architecture are described in Supplementary Table 1, Supplementary Table 2, and Supplementary Fig. 1. As it is standard for U-Nets, the number of dimensions of the input remains unchanged through the layers; the input 3D volume results in a 3D output of dimension $N_i \times N_j \times L$, with $N_i$, $N_j$ and $L$ coming from the image's spatial dimensions, and $L$ also being the pixel latent space size.

For the conditional architecture used for the FLIM dataset, $\boldsymbol{\mathcal{T}}$ is projected to encoding blocks by passing through two learnable fully-connected layers both with 64 nodes, with a GELU activation in-between, and a SiLU activation afterwards (Supplementary Fig. 1). The learned representation then goes through a third fully-connected layer with an output size corresponding to the number of filters of each encoding block (128, 256, 512 and 1024). The values output by this fully-connected layer are then used to scale and shift the input of each encoding block[35].

**Latent space filters**

Algorithm 1 describes the training process of the *Latent Unmixing* method. The training dataset $\boldsymbol{\mathcal{X}}$ contains volumetric data samples $h^i$, along with unmixed ground truth images $x^{i,c}$ for each of $C$ ground truth channels and sample index $i$. At each epoch, the latent representation $z^i$ of a batch of data samples is produced by the 3D U-Net model (line 6). The latent vector is processed using $C$ bandpass filters, resulting in the creation of $C$-channel images where each channel $c$ is the summed signal across filter *bandpass(c)* (lines 8 and 9). The resulting prediction is normalized along the channel dimension so that each pixel of the original image is associated with a probability of belonging to each channel (line 11). The loss is the mean squared error between the predicted ($\hat{x}^{i,c}$) and ground truth ($x^{i,c}$) multi-channel maps of relative abundances (line 13). The loss is backpropagated to update the weights $\theta$ of the network using stochastic gradient descent (line 15). This process is repeated over all batches of training data, and over a number of epochs $e^{max}$.

**Algorithm 1** Training

1: **Input:** $\mathcal{X} : \{(h^{i}_{(N_x \times N_y \times L)}, (x^{i,c}_{(N_x \times N_y)})_{c=1}^{C})\}_{i=1}^{\mathcal{I}}$
2: Initialize randomly U-Net weights $\theta^0$
3: **for** $e = 1, \dots, e^{\max}$ **do**
4:     **for** $i = 1, \dots, \mathcal{I}$ **do**
5:         // Compute latent representation of $h^i$
6:         $z^i \leftarrow \text{U-Net}(h^i \,|\, \theta^e)$
7:         // Apply bandpass filters to $z^i$ and sum
8:         $z^{i,c} \leftarrow \text{bandpass}(z^i, c), \ \forall c$
9:         $\hat{x}^{i,c} \leftarrow \sum_l z_l^{i,c}, \forall c$
10:         // Normalize into a probability of belonging to $c$ (abundance maps)
11:         $\hat{x}^{i,c} \leftarrow \hat{x}^{i,c} \oslash (\sum_c \hat{x}^{i,c})$
12:         // Compute loss $\mathcal{L}^i$
13:         $\mathcal{L}^i \leftarrow \frac{1}{N_x \times N_y} \sum_c \sum_{j,k} (\hat{x}^{i,c}_{j,k} - x^{i,c}_{j,k})^2$
14:         // Update U-Net parameters
15:         $\theta^e \leftarrow \theta^e - \alpha \nabla_\theta \mathcal{L}^i$
16:     **end for**
17: **end for**

For the FLIM dataset, a fifth channel is added to allow the network to assign photons to the background. The first bandpass filter covers the bins 2 to 9, the second 12 to 19, the third 22 to 29, the fourth 32 to 39, and the background channel is created from the remaining bins, i.e. element 1, 10, 11, 20, 21, 30 and 31 (Fig. 3). This allows the network to allocate signal to intersecting elements for pixels with insufficient photons to be accurately classified as any channel other than the background.

**Architecture and training parameters**

Supplementary Table 1 summarizes the architecture parameters for the 3D U-Net (Supplementary Fig. 1) used to obtain the results presented in this paper. All kernel sizes used throughout the network are 3×3×3, and convolutions use zero-padding of size 1 along the three dimensions. Supplementary Table 2 describes the layers inside the DoubleConv3D blocks. For the experiments performed in this study, the receptive field of the final layer spans 110 pixels in width, height, and depth, contrasting with pixelwise approaches for which the predictive capacity is restricted by the very limited information of a single pixel or a narrow and fixed neighborhood.

Down3D blocks are a MaxPooling layer with kernels of shape 2×2×2 and a stride of 2, followed by a DoubleConv3D block. Up3D blocks are DoubleConv3D blocks preceded by a 3D transposed convolution, also with kernels of shape 2×2×2 and a stride of 2. The last convolution, OutConv3D, is a simple convolution with kernels of size 1×1×1 and a stride of 1, to remove a dimension before applying the bandpass filters. Supplementary Table 3 summarizes the training hyper-parameters used for each dataset.

**Generation of the 3D-MNIST dataset**

The 3D-MNIST dataset was created by combining images of up to 3 digits from the original MNIST dataset[33]. To increase the complexity of the unmixing process, we deliberately selected three digits characterized by the highest degree of overlap, ensuring that a substantial number of pixels would count multiple components. To find the trio of digits with the highest fraction of overlap, 1000 samples of each combination of three digits were randomly sampled in the MNIST training set and the average intersection over union was computed. The Intersection Over Union (IOU) for all triplets are shown in Supplementary Fig. 2. The maximum average IOU (0.19) is obtained from the combination of digits 4, 8 and 9.

Following Algorithm 2, we added a third dimension to the images from the original MNIST dataset[33] to simulate a time-resolved measurement in experimental physics. The added signal in the temporal dimension was characterized by an exponential decay function. Each image $i$ was created by randomly sampling up to three images from classes 4, 8 and 9 from the original MNIST training set. Random Poisson noise was added as a fourth target channel to simulate background noise in real experimental physics measurements. The generated dataset was parameterized by a signal-to-noise factor $R$, decay constants for each channel ($\lambda_4$, $\lambda_8$, $\lambda_9$, $\lambda_b$), a time dimension histogram of $L$ bins, and a number of images $I$. The digits were independently randomly rotated, flipped and translated (line 17). Images were normalized to the range [0,1] by dividing each image by its maximum value (line 19). To better simulate noisy experimental conditions, we adjusted the normalized pixel values by sampling a random count for each pixel from a Poisson distribution. The sampling parameter is the product of the pixel value and the Signal-to-Noise Ratio (SNR) factor $R$ (as described in lines 21 and 22). For each count, an associated time value was sampled following an exponential distribution parameterized by the decay constant of the channel (line 24). The time values of each individual pixel are converted to a histogram of $L$ bins of fixed width (line 26), where $L$ is the dimension of the latent space. A background channel was created by randomly sampling values in a Poisson distribution with parameter $\lambda$=5. The resulting average pixel count is 3.9 for the foreground (non-zero) pixels of individual digit channels and 4.7 for the background channel. A time histogram was built following the same method as for the digits channels (lines 29 to 33). There was a probability 0.5 of picking a digit for each of the classes (lines 9 to 13), such that a training sample could contain 0, 1, 2 or 3 digits among 4, 8 and 9. The volumes (4, 8, 9, background) were summed along the channel dimension to create a mixed volume of dimensions ($28 \times 28 \times L$) (line 35). The resulting 3D-MNIST dataset consists of 8,000 mixed volumes generated with $R$=5 and $L$=28, from which 6,400 randomly sampled volumes were used for training and 1,600 for validation. For testing, the same process was repeated using the test MNIST digits to generate 800 test images with overlapping digits and simulated decay.

The SNR (parameter $R=5$ in Algorithm 2) was chosen such that a simple 2D U-Net could not accurately unmix the digits using the intensity signal only (Supplementary Fig. 4), demonstrating that the time dimension is needed to assign individual counts to the right channel. The decay parameters $\lambda_c$ were chosen as 10, 5, 3.33 and 2.5 for digits 4, 8, 9 and the background respectively (Supplementary Fig. 1, top-left).

**Algorithm 2** 3D-MNIST dataset generation

1: **Input:** Signal-to-noise factor $R$
2: **Input:** Exponential distribution parameters $\lambda_4, \lambda_8, \lambda_9, \lambda_b$
3: **Input:** Length of latent space $L$
4: **Input:** Number of images to generate $\mathcal{I}$
5: **Input:** Dataset $\mathcal{X}$ (MNIST)
6: **for** $t = 1, \ldots, \mathcal{I}$ **do**
7:     **for** $c \in \{4, 8, 9\}$ **do**
8:         // Include class $c$ with a probability of 0.5
9:         $b \sim \mathcal{U}(0, 1)$
10:         **if** $b > 0.5$ **then**
11:             $x^{t,c} \leftarrow \varnothing$
12:             Skip to next class $c$
13:         **end if**
14:         // Sampling an image from class $C_c$
15:         $x^{t,c} \leftarrow \text{sample}(\{x^l \in \mathcal{X} \mid y^l = C_c\})$
16:         // Transform the image (rotations, flips, translations)
17:         $x^{t,c} \leftarrow \text{transform}(x^{t,c})$
18:         // Normalize the image to the range [0,1]
19:         $x^{t,c}_{i,j} \leftarrow x^{t,c}_{i,j} / \max_{l,m}(x^{t,c}_{l,m}),\ \forall i, j$
20:         // Generate noisy version of the digit
21:         $u_{i,j} \sim \text{Pois}(R \times x^{t,c}_{i,j}),\ \forall i, j$
22:         $x^{t,c} \leftarrow u$
23:         // Sample $x^{t,c}_{i,j}$ values from exponential distribution
24:         $s^{t,c}_{i,j,k} \sim \text{Exp}(\lambda_c),\ k = 1, \ldots, x^{t,c}_{i,j},\ \forall i, j$
25:         // Convert samples to histogram with $L$ bins
26:         $h^{t,c}_{i,j} \leftarrow \text{hist}(s^{t,c}_{i,j}, L),\ \forall i, j$
27:     **end for**
28:     // Random background noise
29:     $x^{t,b}_{i,j} \sim \text{Pois}(5),\ \forall i, j$
30:     // Sample $x^{t,b}_{i,j}$ values from exponential dist.
31:     $s^{t,b}_{i,j,k} \sim \text{Exp}(\lambda_b),\ k = 1, \ldots, x^{t,b}_{i,j},\ \forall i, j$
32:     // Convert samples to histogram with $L$ bins
33:     $h^{t,b}_{i,j} \leftarrow \text{hist}(s^{t,b}_{i,j}, L),\ \forall i, j$
34:     // Create the joint histogram
35:     $h^t \leftarrow h^{t,4} + h^{t,8} + h^{t,9} + h^{t,b}$
36: **end for**
37: **Return** $\mathcal{X}^G = \{(h^t, (x^{t,c})_{c \in \{4,8,9\}})\}_{t=1}^{\mathcal{I}}$ as generated dataset

## 2D U-Net for unmixing MNIST

To evaluate how the time dimension is exploited by the network to solve the unmixing task, a 2D version of the 3D-MNIST dataset was generated by omitting lines 23-26 and 30-33 in Algorithm 2; that is, by simply not creating the time dimension after creating the noisy overlapping digits. Results obtained by applying a 2D U-Net on this 2D-dataset are shown in Supplementary Fig. 4. The 2D U-Net cannot separate the digits, demonstrating that the third dimension is necessary for unmixing the overlapping digits.

## MLE for unmixing 3D-MNIST

MLE is used as a baseline for the unmixing of 4-channel images with low counts and similar decay constants. MLE uses the limited-memory BFGS algorithm[44] from scipy (version 1.10.1). MLE is used to fit the parameters $A_1$, $A_2$, $A_3$ and $A_4$ in Equation 1 over the 3D-MNIST test set. The parameters are bounded between 0 and 1, and normalized to force all abundance parameters to sum to 1 and respect the second axiom of probability. The decay constants are known and fixed. The resulting confusion matrices, obtained over the 800 test images, are shown in Supplementary Fig. 5.

$$y(t) = \frac{A_1}{0.1} e^{-t/0.1} + \frac{A_2}{0.2} e^{-t/0.2} + \frac{A_3}{0.3} e^{-t/0.3} + \frac{A_4}{0.4} e^{-t/0.4} \quad (1)$$

## Generation of the STED-FLIM dataset

The STED-FLIM training dataset was generated from the *4-channel STED Dataset*[36], which consists of 30 real microscopy images of fixed neurons from primary cell cultures and acute brain slices. Images of different combinations of four neuronal proteins tagged with fluorescent markers were acquired over large fields of view (width between 8 and 16 µm and height between 45 and 65 µm) (Methods). The signal from each fluorescent marker was spectrally separated to obtain ground truth unmixed images with four independent channels. The number of channels for the ground truth dataset is limited by what can be routinely spectrally separated with a standard STED microscope and currently available antibodies (Methods). This is not a limitation of the *Latent Unmixing* method, which can be adapted to any number of target channels.

The images were acquired with a commercial Abberior Expert Line STED microscope. Three different combinations of four proteins of interest were labeled with spectrally discernible fluorescent dyes.

For the first combination, the Microtubule Associated Protein (MAP2) was tagged with ATTO 490LS, the Post-Synaptic Density (PSD-95) with STAR635P, the presynaptic Vesicular Glutamate Transporter (VGLUT1) with Alexa Fluor 488, and the RNA binding protein Fused in Sarcoma (FUS) with Alexa Fluor 594. PSD-95 and FUS were imaged with the STED modality (resolution ~50nm), while MAP2 and VGLUT1 were imaged with the confocal modality (resolution ~200 nm).

For the second combination, hippocampal neurons were transfected with a short-hairpin RNA (shRNA) against βCaMKII and/or with the corresponding resistant construct[45], using Lipofectamine 2000. Transfection was done at 8 Days in Vitro (DIV) and fixation at 12 DIV. The GFP was used as a cytosolic marker, Tau was labeled with ATTO 490LS, the βCaMKII with Alexa Fluor 594, and F-actin with Phalloidin STAR635. βCaMKII and F-actin were imaged with the STED modality, while GFP and Tau were imaged with the confocal modality.

For the third combination, the staining was performed in acute brain slices. PSD-95 was labeled with a nanobody tagged with ATTO 643, the vesicular GABA transporter (VGAT) was labeled with Alexa Fluor 594, VGLUT1 with ATTO 490LS, and the post-synaptic scaffolding protein gephyrin with STAR GREEN. PSD-95, VGAT and VGLUT1 were imaged with the STED modality, while Gephyrin was imaged with the confocal modality.

The protein combinations are summarized in Supplementary Table 4 and the antibodies used in Supplementary Table 5.

Prior to training, the images were cropped into 256 × 256 pixels crops using a sliding window with steps of 128 pixels. Crops were kept in the training set if the mean photon count of each of the 4 channels was at least 1 photon per pixel. This operation resulted in a set of 1132 crops for training. To augment the data during training, samples were further randomly cropped to a size of 64 × 64 pixels, randomly flipped and rotated, channels were permuted, and up to two channels were removed. The time dimension was created during training by first sampling one lifetime value $\tau_1 \sim \mathcal{N}(1.5,0.5)$ for the first non-zero channel, then adding $\Delta \sim \Gamma(1,0.5)$ for each subsequent non-zero channel. The minimal distance between lifetime values was set at 0.2 nanoseconds, so any value less than 0.2 was changed to 0.2. Random arrival time values for each photon are sampled in an exponential decay curve parameterized by the channel's lifetime. We chose to pick lifetime values in the range from 0 to 6 nanoseconds because this range includes the majority of the commercially available dyes[2,6]. Training a model on a predefined set of lifetime values would not capture the true measured distributions of fluorescence lifetime in a real sample, especially for live-cell FLIM.

To generate the *Confocal- and STED-FLIM images of neuronal proteins dataset*[37], which consists of real FLIM images, we equipped the microscope with a Time Tagger Ultra 8 from Swabian Instruments GmbH which enabled Time-Correlated Single Photon Counting (TCSPC) measurements for 2 imaging channels. The red imaging channel (605-625 nm) using a 561 nm excitation laser and the far-red imaging channel (650-720 nm) using a 640 nm excitation laser. To acquire STED-FLIM images, a 775 nm depletion laser was used for both the red and far-red channels.

We acquired Confocal-FLIM (Figure 6a-b) and STED-FLIM (Figure 6c, left) images of the pre-synaptic protein Bassoon tagged with CF594 and the post-synaptic protein Homer1c tagged with STAR ORANGE in fixed neuronal cultures. These two fluorophores have similar excitation and emission spectra and were both imaged using the red imaging channel. STED-FLIM images of the cytoskeletal protein alpha-Tubulin tagged with Alexa Fluor 647 and Bassoon tagged with STAR 635 were acquired in fixed neuronal cultures using the far-red channel. In live neuronal cultures, SiR-Actin and Tubulin-LIVE 610 were used to tag F-actin and Tubulin respectively and they were both imaged using the far-red imaging channel. Further image acquisition details for the FLIM experiments can be found in Supplementary Table 8.

**Acquisition of microscopy images**

<u>Sample preparation</u>

Fluorescence microscopy images from both neuronal cell cultures (Fig. 4 and Supplementary Fig. 8) and acute brain slices (Supplementary Fig. 9) were included to the datasets used in this study.

Primary rat neuronal cell cultures were prepared following established protocols[46] in accordance with procedures approved by the animal care committee of Université Laval. Following dissociation from dissected cortex and hippocampus of P0-P1 postnatal rats, cells were plated onto PDL-Laminin coated glass coverslips (12 mm) in a 24-well plate at a density of 100k cells/well. Cells were fixed for 20 minutes with a 4% PFA solution following an established protocol[47] at 12 to 15 DIV.

Acute mouse brain slices were prepared using an optimized version of an established protocol[48] in accordance with the procedures approved by the animal care committee of Université Laval. Coronal slices of prefrontal cortex and hippocampus were cut at 400 µm and kept at 37 degrees Celsius in a slicing solution containing (in mM) 93 NMDG, 20 HEPES, 2.5 KCl, 5 Sodium Ascorbate, 2 Thiourea, 3 Sodium Pyruvate, 0.5 $CaCl_2$, 10 $MgSO_4$, 1.2 NaH2PO4, 30 $NaHCO_3$, and 25 glucose, with a pH adjusted between 7.50 - 7.55 and an osmolarity between 300 – 310 mOsm, saturated with 95% O2 and 5% $CO_2$ for 15 minutes. Slices were transferred at room temperature in Holding Solution containing (in mM) 92 NaCl, 20 HEPES, 2.5 KCl, 1.2 $NaH_2PO_4$, 25 Glucose, 30 $NaHCO_3$, 5 Sodium Ascorbate, 2 Thiourea, 3 Sodium Pyruvate, 2 $MgSO_4$ and 2 $CaCl_2$, with a pH adjusted between 7.50 - 7.55 and an osmolarity between 300 – 310 mOsm, saturated with 95% $O_2$ and 5% $CO_2$. Slices were fixed with 3% glyoxal / 0.8% acetic acid for 24 hours at 4 degrees Celsius.

After fixation, the samples underwent three 5-minute washes in phosphate-buffered saline (PBS) supplemented with 100 mM glycine. Subsequently, the cells were permeabilized and blocked with a blocking solution composed of 0.1% Triton X-100 and 2% goat serum (GS) in PBS for 30 minutes before proceeding to immunostaining. Immunostaining involved incubation with primary antibodies (PAB) and secondary antibodies (SAB) diluted in the blocking solution at room temperature (RT, Supplementary Table 5). PAB were incubated for 2 hours (24 hours for the brain slices), followed by three PBS washes. SAB were then incubated for 1 hour (2 hours for the brain slices) and subsequently washed three times with PBS. In the case of F-actin labelling, phalloidin fused with STAR635 and diluted in blocking solution was incubated for 2 additional hours at RT, and subsequently washed three times with PBS. Finally, coverslips and brain slices were mounted onto microscopy slides using Polyvinyl alcohol mounting medium with DABCO (Sigma, cat:10981).

For live-cell imaging experiments, cultured hippocampal neurons were incubated in culture media containing 0.5 μM Tubulin-LIVE610 (Abberior, cat: LV610-0141) for 1h at 37°C . Prior to imaging, the coverslips were transferred to a solution of artificial cerebrospinal fluid (aCSF) containing 0.5μM SiR-Actin (Spirochrome, cat:SC001) for 15 min. aCSF contains (in mM) 98 NaCl, 5 KCl, 10 HEPES, 5 MgCl, 0.6 $CaCl_2$, 10 Glucose with a pH adjusted to 7.4 and an osmolarity of 239 mOsm. Imaging in aCSF was performed using a gravity perfusion system. Detailed information and further details for the imaging acquisition parameters can be found in Supplementary Table 6 and Supplementary Table 7.

## MLE for FLIM

MLE was tested over pixel counts up to 500 to assess the number of photons required to correctly estimate the relative abundance $A$, in the case where there are two emitters in the pixel and their lifetime values are known (equation 2). The absolute error between the proportion of the first component ($A$) predicted by MLE and the actual value is shown in Supplementary Fig. 6 for varying numbers of photons, ground truth values of $A$, and combinations of lifetime values $\tau_1$ and $\tau_2$. MLE was applied here by bounding the value of $A$ between 0 and 1, and the initial guesses are values between 0 and 1 randomly picked at each iteration. MLE shows lower performance mostly when photon counts are low and the proportion of both classes is unbalanced ($A \neq 0.5$).

$$y(t) = \frac{A}{\tau_1} e^{-t/\tau_1} + \frac{(1-A)}{\tau_2} e^{-t/\tau_2} \quad (2)$$

## Phasor analysis method for FLIM

We used an established method[49] to unmix the FLIM images with the phasor analysis method. We computed the reference positions of each channel using the same control images as for computing the average lifetime for the *Latent Unmixing* method, presented in Supplementary Table 9. For the simulated FLIM test sets, controls were generated from a uniform field of view of 300 × 300 pixels with 300 photons per pixels and randomly sampled lifetimes. The proportion of photons assigned per pixel to each channel is derived from a linear interpolation between the centroid positions of the phasor clusters obtained from the controls (Supplementary Fig. 7). We used the first and second harmonics of the Fourier transform to project the decay curves to the phasor space.

### Latent unmixing for real STED-FLIM measurements

For real STED images, vector $\boldsymbol{\mathcal{T}}$ has three non-negative values (Supplementary Table 10): the mean of the periphery lifetime of both emitters[39], the center lifetime of the first emitter, and the center lifetime of the second emitter. Shown in Fig. 6 are the photons associated with the central PSF of both emitters.

### Unconditional latent unmixing for FLIM

A version of the *Latent Unmixing* model was trained for two-channel unmixing without conditioning. The lifetime values in training were fixed at 2.77 and 3.55 ns for the Bassoon and Homer1c channels, respectively. This model could make accurate predictions only if the lifetime values of the test data for both channels was exactly 2.77 and 3.55 ns (Supplementary Fig. 11), rendering the model useless for any biological context where slight deviations in the lifetime are to be expected. This result led us to develop the conditional *Latent Unmixing* model for FLIM data, allowing better generalizability over the lifetime values.

### STED-FLIM Test sets

To test the *Latent Unmixing* method for 4-channel unmixing, a test set was built from the 12 test images included in the *4-channel STED Dataset*[36]. Two combinations of lifetime values were used: $\boldsymbol{\mathcal{T}}$ = [0.2, 2.1, 4.0, 5.9] and $\boldsymbol{\mathcal{T}}$ = [1.0, 2.0, 3.0, 4.0]. The first combination was chosen because they are the furthest apart lifetime values that can be defined in the [0, 6] ns range. The second combination was chosen as an easy case that could be achieved experimentally, with each lifetime 1 ns apart from the others. We used well separated lifetime values to improve the phasor analysis method's performance, but even with the most separated values, 4-channel unmixing was still not achievable with this method. This test set counts 24 4-channel images, or 96 single-channel images for computing performance metrics. For 3-channel unmixing, the same 12 intensity images were used to build the test set using the same lifetime combinations ($\boldsymbol{\mathcal{T}}$ = [0.2, 2.1, 4.0, 5.9] and $\boldsymbol{\mathcal{T}}$ = [1.0, 2.0, 3.0, 4.0]). We used all possible combinations of 3 channels (and 3 corresponding lifetime values) from the 4 total channels, bringing the total number of 3-channel test images to 96, or 288 single-color channels. For 2-channel unmixing, all combinations of 2 channels (and corresponding lifetime values) were used, resulting in 144 2-channel images, or 288 single-channel images. This is the test set that was used for the results of figure Fig. 3**e,f**.

To test the limits in terms of the proximity between lifetime values, a 2-channel test set was generated from the same 12 intensity test images. The lifetime of the second channel is fixed at 3 ns for all generated FLIM images. The lifetime of the first channel was changed from 0.2 ns to 2.8 in increments of 0.2 ns, resulting in 168 2-channel test images or 336 single-channel images for computing the metrics presented in Fig. 5.

**Generation of the $S^2$-MMF dataset**

A vector finite difference mode-solver[50] was used to find mode properties around 1550 nm for various index profiles and fiber geometries. All fibers simulated in the dataset support up to 6 modes. The mode spatial distribution and effective index dependence on wavelength (dispersion) were taken into account. For all data samples, we considered a spectral range of 6 nm for a fiber length of 5 meters. The dataset includes simulated measurements with various combinations of up to 6 propagating modes, different fiber geometries (step-index with varying numerical apertures and core size, GRIN, W-type, H-type and a realistic step-index geometry with a central dip), and different resulting mode-specific intensities ($F$). We used 1500 randomly generated mode combinations to train the network and 175 for testing.

**Mode decomposition in multi-mode fibers**

Supplementary Fig. 13 decomposes the data points of Fig. 7**e** by mode order, also including 0-intensity target modes. Mode decomposition performance does not significantly vary for different spatial distributions, as all modes tend to follow a similar trend towards the diagonal, with a few outliers.

In addition to the correlation between the global relative intensity of each mode, the *Latent Unmixing* method also correctly uncovers their spatial distributions (Supplementary Fig. 14). The predicted pixel-wise intensity tends to be under-estimated compared to the ground truth intensity (slope of the regression >1, solid white line in Supplementary Fig. 14). This is again a result of the model tending to assign some relative intensity to non-propagating modes (non-zero elements where true intensity = 0 in Supplementary Fig. 14). Still, the pixel-wise Pearson correlation coefficient averaged over the whole test set is >0.77 for all 6 modes. Supplementary Fig. 15 shows qualitative results over three different fiber geometries from the test set.

**Statistical analysis**

For statistical analysis, the normality of each distribution is verified with the Shapiro-Wilk test with $\alpha=0.05$. When both distributions are normal, the null hypothesis that the two groups have the same variance is verified with the F-test with $\alpha=0.05$. If both variances are equal, the null hypothesis that the two groups have the same average is verified with the T-test for the means of two independent samples. If both variances cannot be assumed equal, the same null hypothesis is verified with the Welch's T-test. If at least one of the two distributions is not normal, the null hypothesis is verified with the non-parametric one-way ANOVA test.

For the results presented in Fig. 5**c**, the null hypothesis of normality could be rejected for all distributions based on the Shapiro-Wilk test. The null hypothesis that the two distributions have the same population mean was verified with the one-way ANOVA test. For the results presented in Fig. 5**d**, the null hypothesis of normality could be rejected for all distributions based on the Shapiro-Wilk test. The null hypothesis that the two distributions of samples come from populations with the same median was verified with the non-parametric Mood’s median test.

**Acknowledgements**

Albert Michaud-Gagnon and Marie Lafontaine for the implementation of the 2-species phasor analysis unmixing method.

**DECLARATIONS**

**Funding**

Funding was provided by grants from the Natural Sciences and Engineering Research Council of Canada (NSERC) (RGPIN-06704-2019 to F.L.C. and RGPIN-2019-06706 to C.G.), Fonds de Recherche Nature et Technologie (FRQNT) Team Grant (2021-PR-284335 to F.L.C and C.G.), the New Frontiers in Research Fund (NFRF) Exploration Grant (NFRFE-2020-00933 to F.L.C.), the Canadian Institute for Health Research (CIHR) (202109PJT-471107-NSB-CFBA-12805 to F.L.C.), the Sentinelle Nord Initiative (F.L.C. and C.G.), and the Neuronex Initiative (National Science Foundation 2014862, Fond de recherche du Québec - Santé 295824) (F.L.C.). C.G. is a CIFAR Canada AI Chair and F.L.C. is a Canada Research Chair Tier II. C.B. is supported by scholarships from NSERC, from the FRQNT (https://doi.org/10.69777/301611), from the FRQNT strategic cluster UNIQUE, and by a Leadership and Scientific Engagement Award from Université Laval. V.B. is supported by a scholarship from NSERC.

**Competing interests**

The authors have no competing interests to declare that are relevant to the content of this article.

**Code availability**

All scripts used to generate the 3D-MNIST dataset and the time dimension of the STED-FLIM dataset, train the models, test the models with trained weights, and produce the results presented here is available to download at https://github.com/FLClab/LatentUnmixing. All experiments were conducted using Python 3.8.10, with the library versions indicated in the Github repository.

**Data availability**

All datasets are available to download at https://s3.valeria.science/flclab-unmixing/index.html. The *4-channel STED Dataset* is available at 10.5281/zenodo.15447483 and the *Confocal- and STED-*

*FLIM images of neuronal proteins Dataset* is available at 10.5281/zenodo.15438495.

**Authors contribution**

C.B., F.L.C. and C.G. have conceptualized the method. C.B. has implemented the method and ran the experiments. A.D. has implemented the phasor analysis method and acquired the real FLIM images. V.B. has generated the simulated $S^2$ data and analyzed the results. J-M.B., J.C. and A.P.R. have acquired the 4-colors confocal and STED images used for training and testing. C.B., V.B., C.G. and F.L.C. have written the manuscript.

# Supplementary Material

| Block | Type | In Channels | Out Channels |
| --- | --- | --- | --- |
| 1 | DoubleConv3D | $L$ | 64 |
| 2 | Down3D | 64 | 128 |
| 3 | Down3D | 128 | 256 |
| 4 | Down3D | 256 | 512 |
| 5 | Down3D | 512 | 1024 |
| 6 | Up3D | 1024 | 512 |
| 7 | Up3D | 512 | 256 |
| 8 | Up3D | 256 | 128 |
| 9 | Up3D | 128 | 64 |
| 10 | OutConv3D | 64 | 1 |

**Supplementary Table 1 -** Architecture of the 3D-UNet used to produce the results presented in this paper. Dimension $L$ is 28 for 3D-MNIST, 40 for STED-FLIM, and 60 for $S^2$-MMF. Blocks 5 and 6 were removed for $S^2$-MMF.

| Layer | Type | Channels |
| --- | --- | --- |
| 1 | Convolution | In Ch. → Out Ch. |
| 2 | BatchNorm3D | Out Ch. → Out Ch. |
| 3 | ReLU | |
| 4 | Convolution | Out Ch. → Out Ch. |
| 5 | BatchNorm3D | Out Ch. → Out Ch. |
| 6 | ReLU | |

**Supplementary Table 2 -** Layers of the DoubleConv3D block.

| **Hyperparameter** | **3D-MNIST** | **STED-FLIM** | **$S^2$-MMF** |
|---|---|---|---|
| Learning rate | 0.01 | 0.01 | 0.01 |
| Batch size | 64 | 12 | 16 |
| Input size | $28^2$ x 28 | $96^2$ x 40 | $64^2$ x 60 |
| Output size | $28^2$ x 28 | $96^2$ x 40 | $64^2$ x 60 |
| Number of filters | 4 | 5 | 6 |
| Filter width | 7 | 8 | 10 |
| Number of images | 8,000 | 1,132 | 1,500 |
| Best epoch | 167 | 272 | 182 |

**Supplementary Table 3 -** Training hyperparameters used to obtain the results presented in this paper. Best epoch is the epoch where the validation loss reaches a minimum.

| **Number of training images** | **Number of testing images** | **Protein 1** | **Protein 2** | **Protein 3** | **Protein 4** |
|---|---|---|---|---|---|
| 10 | 4 | VGLUT1<br>AF488<br>Confocal | MAP2<br>ATTO 490LS<br>Confocal | FUS<br>AF594<br>STED | PSD95<br>STAR635P<br>STED |
| 10 | 4 | GFP<br>-<br>Confocal | Tau<br>ATTO 490LS<br>Confocal | βCaMKII<br>AF594<br>STED | Actin<br>STAR635<br>STED |
| 10 | 4 | Gephyrin<br>STAR GREEN<br>Confocal | VGLUT1<br>AF488<br>STED | VGATr<br>AF594<br>STED | PSD95<br>STAR635P<br>STED |

**Supplementary Table 4 -** Composition of the 42 intensity images from the *4-channel STED dataset* used for building the training and test sets for STED-FLIM.

| Primary Antibodies | | | |
|---|---|---|---|
| **Antibody** | **Supplier** | **Catalogue no.** | **Dilution** |
| Mouse anti-Bassoon | Enzo | ADI-VAM-PS003 | 1 : 500 |
| Rabbit anti-Homer1 | Synaptic Systems | 160003 | 1 : 500 |
| Mouse anti-MAP2 | Sigma-Aldrich | M4403 | 1 : 1000 |
| Mouse anti-βCaMKII | Invitrogen | 13-9800 | 1 : 100 |
| Rabbit anti-FUS | Sigma-Aldrich | HPA008784 | 1 : 500 |
| Guinea Pig anti-VGLUT1 | Sigma-Aldrich | Ab5905 | 1 : 500 |
| Guinea Pig anti-Tau | Synaptic Systems | 314004 | 1 : 250 |
| Fluotag X2 anti-PSD95 STAR 635P | Nanotag Biotechnologies | N3702-Ab635P-L | 1 : 500 |
| Fluotag X2 anti-PSD95 ATTO 643 | Nanotag Biotechnologies | N3702-At643-L | 1 : 500 |
| Mouse anti-Gephyrin | Synaptic Systems | 147 021 | 1 : 500 |
| Rabbit anti-VGAT | Synaptic Systems | 131003 | 1 : 500 |
| Mouse anti-Tubulin | Sigma-Aldrich | T5168 | 1 : 500 |
| Rabbit anti-Bassoon | Synaptic Systems | 141003 | 1 : 500 |
| **Secondary Antibodies** | | | |
| **Antibody** | **Supplier** | **Catalogue no.** | **Dilution** |
| GAM CF594 | Sigma-Aldrich | SAB4600321 | 1 : 500 |
| GAR STAR ORANGE | Abberior | STORANGE-1002 | 1 : 250 |
| GAM ATTO490LS | Hypermol | 2109 | 1 : 250 |
| GAGP ATTO490LS | Hypermol | 2704 | 1 : 250 |
| GAGP Alexa Fluor 488 | Sigma-Aldrich | A-11073 | 1 : 250 |
| GAR Alexa Fluor 594 | Invitrogen | A32740 | 1 : 100 |
| GAM Alexa Fluor 594 Plus | Thermofisher | A32742 | 1 : 100 |
| GAM STAR GREEN | Abberior | STGREEN-1001 | 1 : 250 |
| GAGP ATTO 490LS | Hypermol | 2704 | 1 : 250 |
| (Dye) Phalloidin STAR 635 | Sigma Aldrich | 30972 | 1 : 50 |
| GAR STAR 635P | Abberior | ST635P-001 | 1 : 250 |
| GAM Alexa Fluor 647 | Jackson Immunoresearch | 115-605-166 | 1 : 250 |

**Supplementary Table 5 -** Antibodies used for immunostaining.

| **Parameter** | **Confocal AF 488** | **Confocal Atto 490LS** | **STED AF 594** | **STED STAR 635P** |
|---|---|---|---|---|
| **Excitation wavelength** | 485 nm | 485 nm | 561 nm | 640 nm |
| **Excitation power** | 6.6 μW | 6.6 μW | 3.8 μW | 8.4 μW |
| **Depletion wavelength** | | | 775 nm | 775 nm |
| **Depletion power** | | | 156 mW | 156 mW |
| **Detection wavelengths** | 500-550 nm | 650-720 nm | 605-625 nm | 650-720 nm |
| **Pixel dwelltime** | 10 μs | 10 μs | 10 μs | 10 μs |
| **Line steps** | 2 | 2 | 10 | 6 |
| **Detection gating delay** | | | 750 ps | 750 ps |
| **Detection gating width** | | | 8 ns | 8 ns |
| **Pixel size** | 20 nm | 20 nm | 20 nm | 20 nm |
| **Pinhole size** | 1.0 A.U. | 1.0 A.U. | 1.0 A.U. | 1.0 A.U. |

**Supplementary Table 6 -** Imaging parameters for acquisition of 4-color images (cultures).

| **Parameter** | **Confocal Star Green** | **STED Atto 490LS** | **STED AF 594** | **STED ATTO 643** |
|---|---|---|---|---|
| **Excitation wavelength** | 485 nm | 485 nm | 561 nm | 640 nm |
| **Excitation power** | 7.7 μW | 7.7 μW | 3.25 μW | 5.0 μW |
| **Depletion wavelength** | | 775 nm | 775 nm | 775 nm |
| **Depletion power** | | 132 mW | 220 mW | 132 mW |
| **Detection wavelengths** | 500-550 nm | 650-720 nm | 605-625 nm | 650-720 nm |
| **Pixel dwelltime** | 15 μs | 15 μs | 15 μs | 15 μs |
| **Line steps** | 6 | 10 | 8 | 4 |
| **Detection gating delay** | | 750 ps | 750 ps | 750 ps |
| **Detection gating width** | | 8 ns | 8 ns | 8 ns |
| **Pixel size** | 20 nm | 20 nm | 20 nm | 20 nm |
| **Pinhole size** | 1.0 A.U. | 1.0 A.U. | 1.0 A.U. | 1.0 A.U. |

**Supplementary Table 7 -** Imaging parameters for acquisition of 4-color images (slices).

| | Confocal-FLIM | STED-FLIM | | |
|---|---|---|---|---|
| **Parameter** | Homer Bassoon | Homer Bassoon | Tubulin Bassoon | Actin-Tubulin Live-cell |
| **Excitation wavelength** | 561 nm | 561 nm | 640 nm | 640 nm |
| **Excitation power** | 2.6 μW | 2.5 μW | 2.0 μW | 3.6 μW |
| **Depletion wavelength** | | 775 nm | 775 nm | 775 nm |
| **Depletion power** | | 44, 132 mW | 22, 44 mW | 44 mW |
| **Detection wavelengths** | 605-625 nm | 605-625 nm | 650-720 nm | 650-720 nm |
| **Pixel dwelltime** | 15 μs | 12 μs | 15 μs | 15 μs |
| **Line steps** | 4 | 38 | 10 | 16 |
| **Detection time bins** | 250 bins of 80 ps, 20 ns range | 250 bins of 80 ps, 20 ns range | 250 bins of 80 ps, 20 ns range | 250 bins of 80 ps, 20 ns range |
| **Pixel size** | 20 nm | 20 nm | 20 nm | 20 nm |
| **Pinhole size** | 1.0 A.U. | 1.0 A.U. | 1.0 A.U. | 1.0 A.U. |

**Supplementary Table 8 -** Imaging parameters for acquisition of real FLIM images.

| Protein | Fluorophore | # images | Average (counts/image) | Estimated Lifetime |
|---|---|---|---|---|
| Homer1c | STAR ORANGE | 12 | $1.8 \times 10^5$ | 3.30 ± 0.01 ns |
| Bassoon | Cf 594 | 12 | $3.3 \times 10^5$ | 2.40 ± 0.01 ns |

**Supplementary Table 9 -** Lifetime parameter (mean ± standard deviation) computed from fitting a mono-exponential curve with MLE over control samples with a single tagged protein, imaged with confocal microscopy.

| Protein | Fluorophore | # images | Average (counts/image) | Estimated Lifetime (Periphery) | Estimated Lifetime (Center) |
|---|---|---|---|---|---|
| Homer1c | STAR ORANGE | 20 | $8.3 \times 10^5$ | 0.44 ± 0.06 ns | 3.31 ± 0.02 ns |
| Bassoon | CF 594 | 18 | $3.4 \times 10^5$ | 0.39 ± 0.06 ns | 2.88 ± 0.07 ns |
| αTubulin | Alexa Fluor 647 | 31 | $8.2 \times 10^6$ | 0.39 ± 0.04 ns | 1.87 ± 0.06 ns |
| Bassoon | STAR 635P | 51 | $1.7 \times 10^6$ | 0.44 ± 0.12 ns | 3.59 ± 0.26 ns |
| Tubulin | LIVE 610 | 54 | $2.2 \times 10^6$ | 0.45 ± 0.06 ns | 3.74 ± 0.07 ns |
| Actin | SiR | 54 | $2.5 \times 10^5$ | 0.43 ± 0.06 ns | 3.39 ± 0.23 ns |

**Supplementary Table 10 -** Lifetime parameters (mean ± standard deviation) computed from fitting a bi-exponential curve with MLE over control samples with a single tagged protein, imaged with STED microscopy.

| Mode index | Pearson ρ |
|---|---|
| Mode 1 | 0.869 |
| Mode 2 | 0.778 |
| Mode 3 | 0.822 |
| Mode 4 | 0.813 |
| Mode 5 | 0.801 |
| Mode 6 | 0.918 |

**Supplementary Table 11 -** Pearson correlation coefficient of the relative intensity of each pixel computed over the test set (N=175 simulated measurements, 716,800 pixels).

| Number of modes | Pearson ρ |
|---|---|
| 1 mode | 1.000 |
| 2 modes | 0.781 |
| 3 modes | 0.870 |
| 4 modes | 0.970 |
| 5 modes | 0.947 |
| 6 modes | 0.675 |

**Supplementary Table 12 -** Pearson correlation on the relative intensity of the modes *F* obtained on the test set for different numbers of propagating modes (N=175 total simulated measurements).

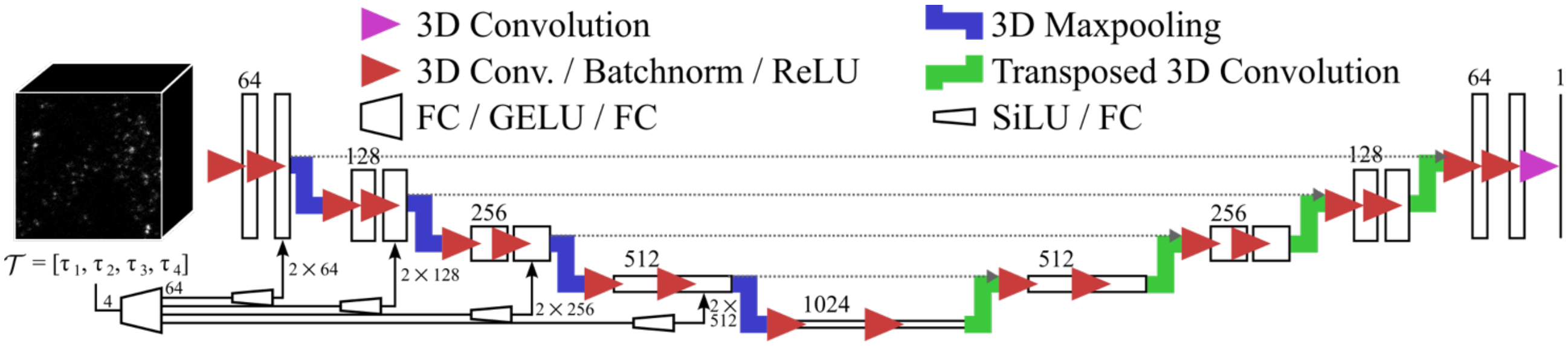

**Supplementary Figure 1 -** The 3D U-Net network architecture used for unmixing the STED-FLIM dataset. For 3D-MNIST and $S^2$-MMF, the conditioning branch with the encoder and layer-specific fully-connected layers is not used. A smaller version is used for $S^2$-MMF, with 3 Down and Up blocks instead of 4.

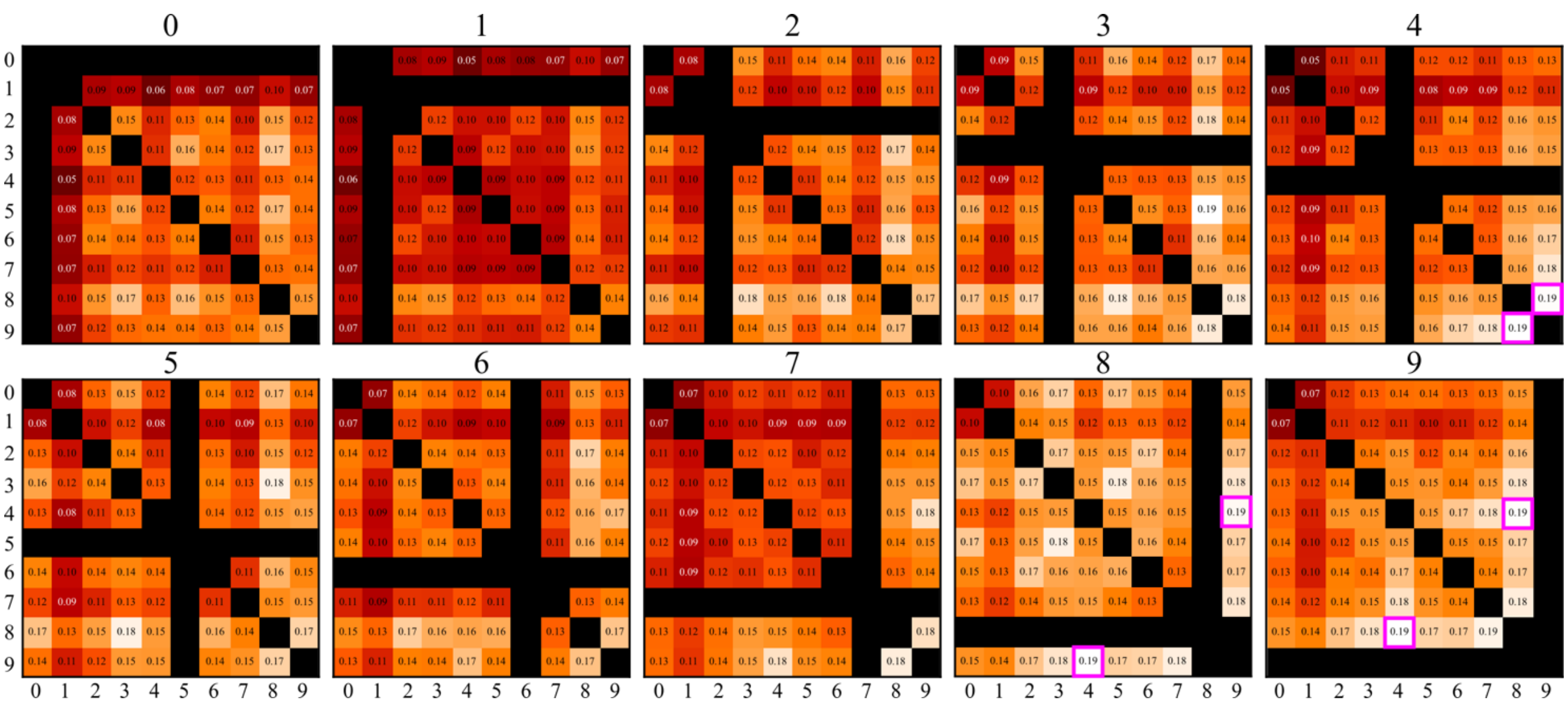

**Supplementary Figure 2 -** Intersection over union between all digits of MNIST (from top-left to bottom-right: 0, 1, ... 9). Diagonals are left blank to avoid identical pairs of digits. Digits 4, 8, and 9 (highlighted in magenta) have the highest average intersection over union (0.19), making it the most difficult trio of digits to separate.

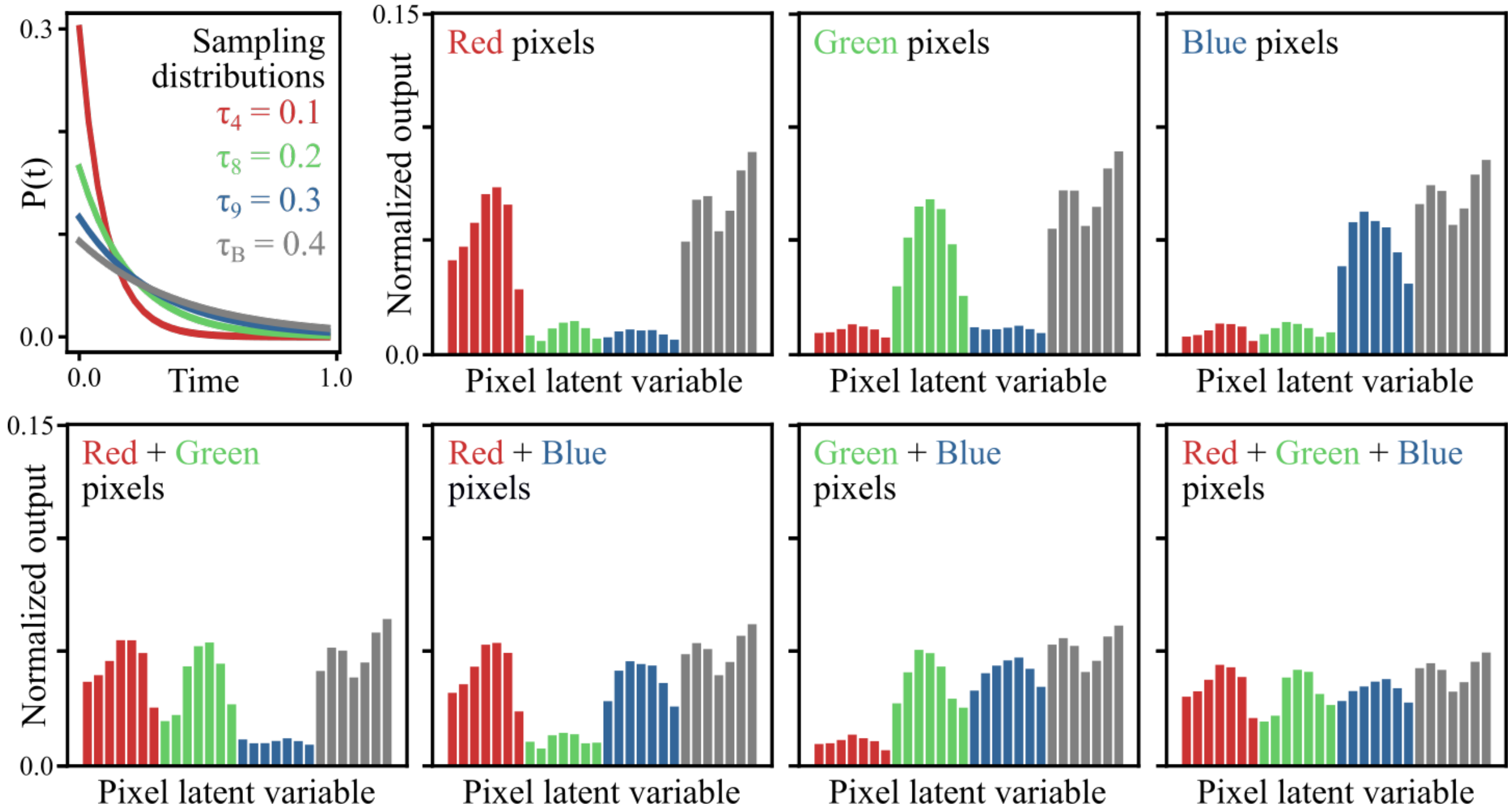

**Supplementary Figure 3 -** Top-left: sampling distributions for three-color MNIST with exponential decay parameters $\lambda_c = 1/\tau_c$ of 10, 5, 3.33 and 2.5 for digits 4 (red), 8 (green), 9 (blue), and the background (gray), respectively. Other subfigures: the pixel output value of the pixel latent variable, averaged over all pixels from the simulated test set. Each subfigure represents a different combination of ground truth pixel composition. All pixels contain background noise (gray bins).

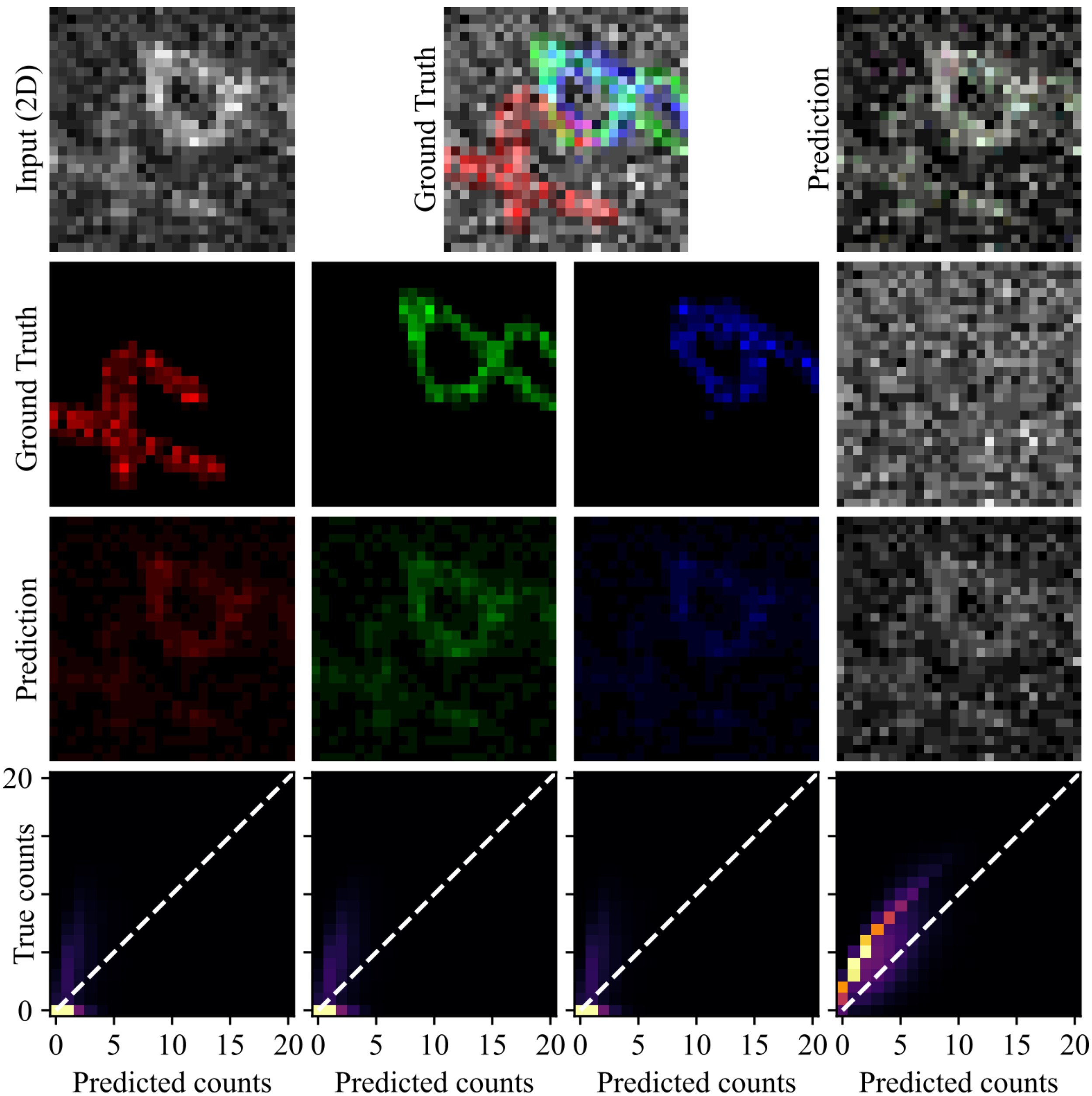

**Supplementary Figure 4 -** Example result obtained with a 2D U-Net applied for unmixing the 2D version of the noisy overlapping MNIST digits. All channels are color scaled to the maximum of the corresponding target channel.

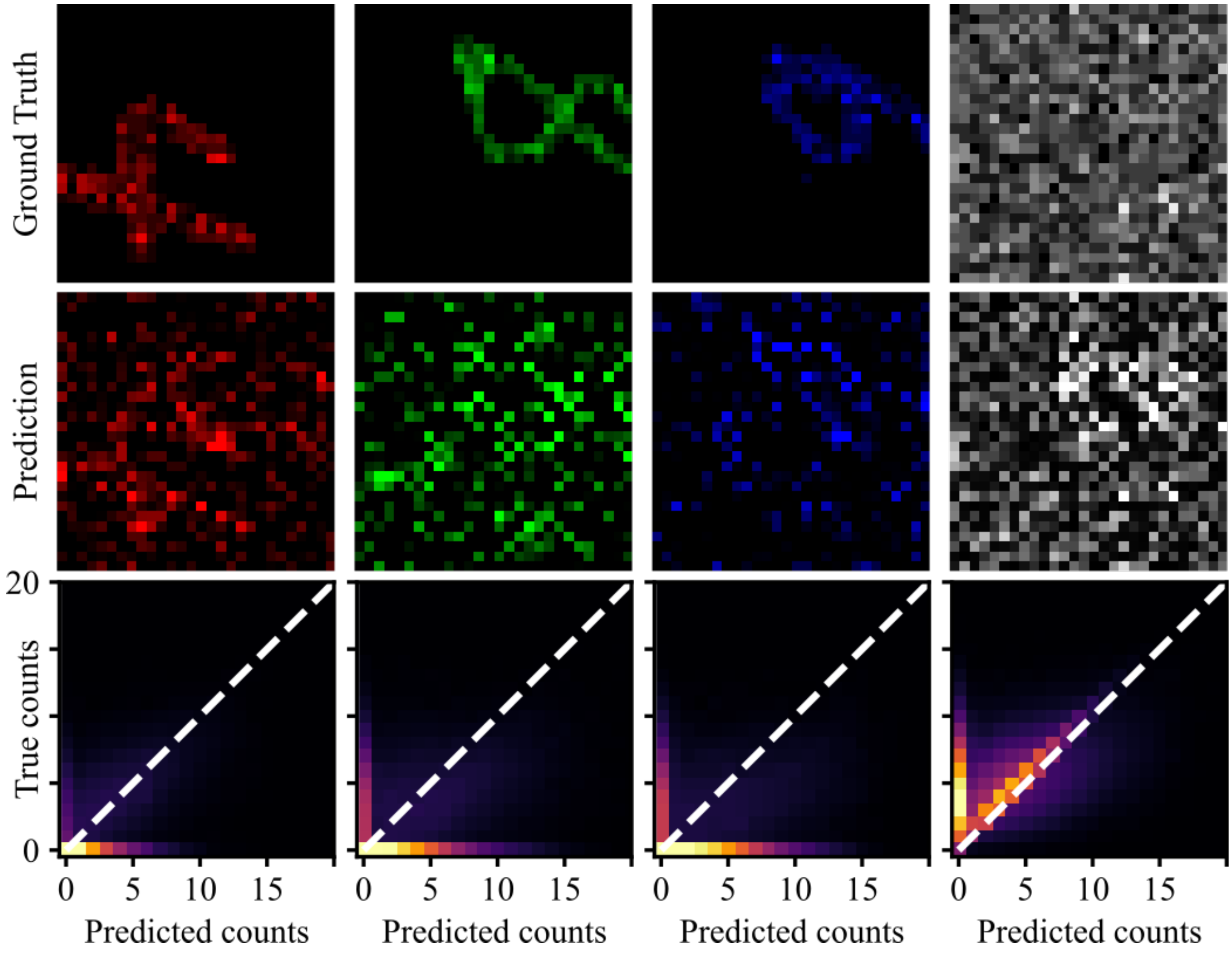

**Supplementary Figure 5 -** Channel-wise results on the test set, with ground truth (top), MLE prediction with 4 abundance parameters to adjust (eq. 1, middle) and correlation matrices between true and predicted photons over all pixels from the 3D-MNIST test set (N=800 images, 627,200 pixels).

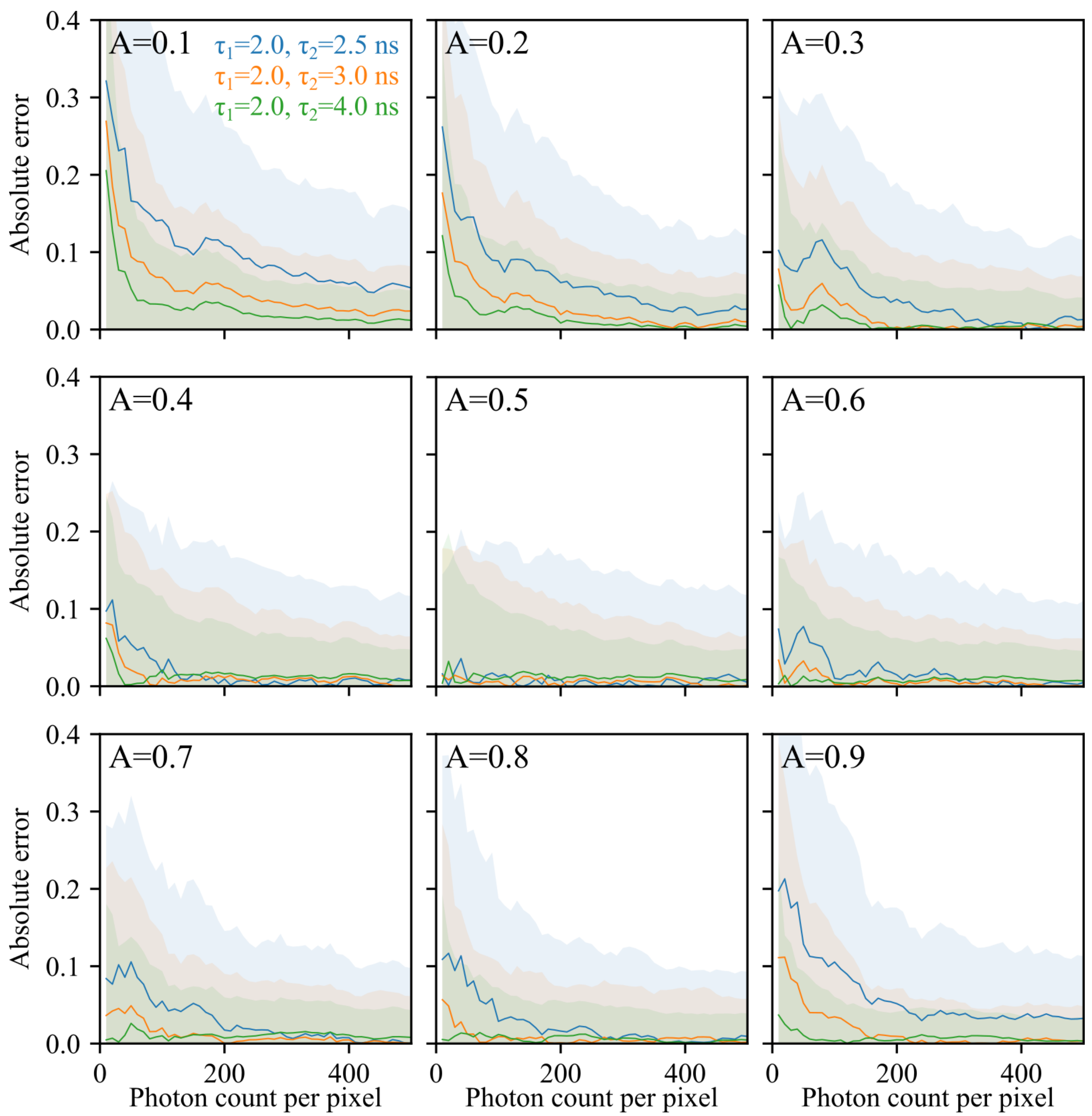

**Supplementary Figure 6 -** Absolute error ($| A_{true} - A_{predicted} |$) between actual and predicted value using MLE for abundance parameter $A$ in eq. 2 for a single simulated pixel. For each graph, target proportion $A$ is incremented by 0.1. Lines show the average of 100 repetitions with different random seeds and initial values, and the shaded region shows the standard deviation.

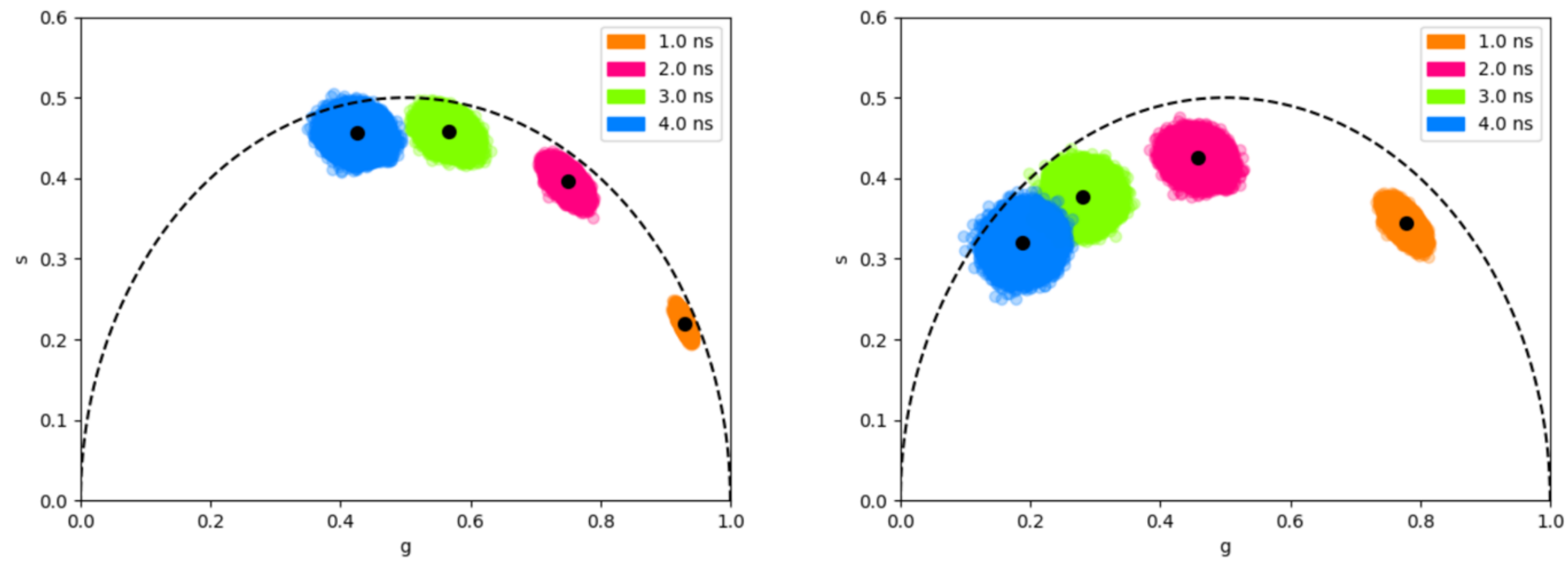

**Supplementary Figure 7 -** Phasor plots obtained from simulated controls with 90,000 pixels per channel and 300 photons per pixels. Left and right graphs show first and second harmonics. Each point is a pixel, color-coded by channel. Black points are the centroids of all points belonging to each channel.

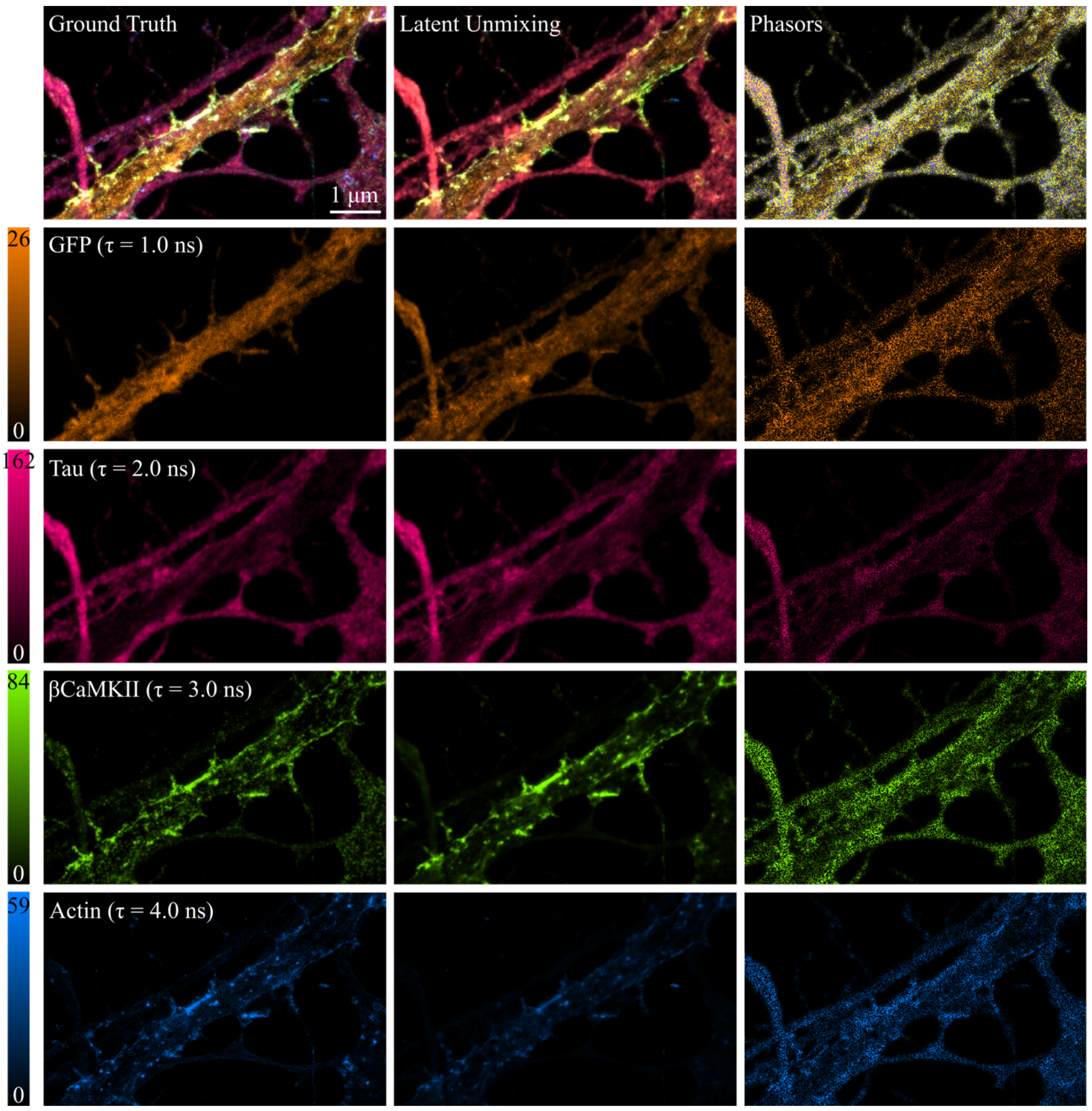

**Supplementary Figure 8 -** Image from the STED-FLIM simulated test set for the unmixing of 4 channels with the *Latent Unmixing* and the phasors analysis methods. All colormaps are scaled from 0 to the 99.9th percentile of the ground truth.

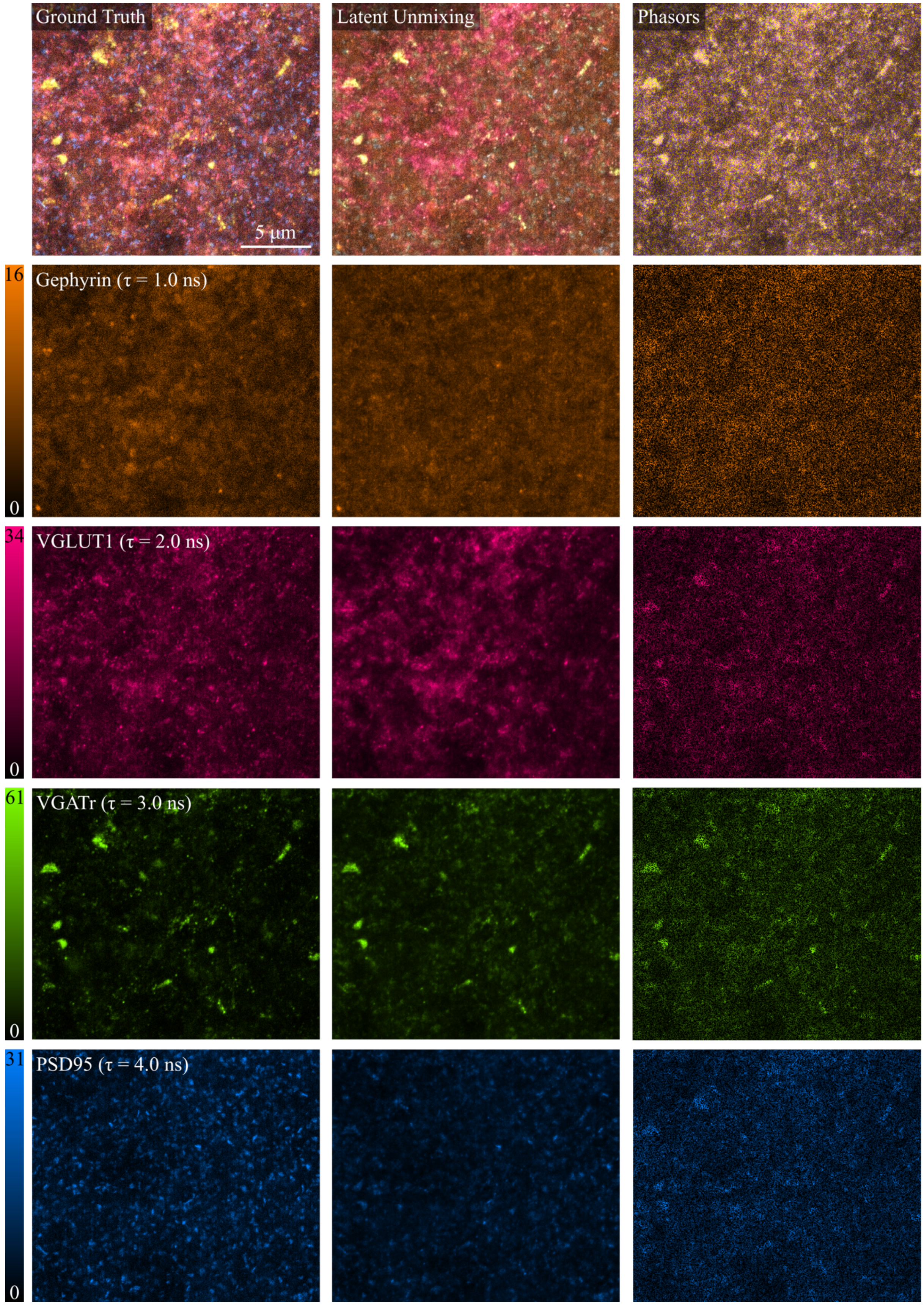

**Supplementary Figure 9 -** Image from the STED-FLIM simulated test set for the unmixing of 4 channels with the *Latent Unmixing* and the phasors analysis methods. All colormaps are scaled from 0 to the 99.9th percentile of the ground truth channels.

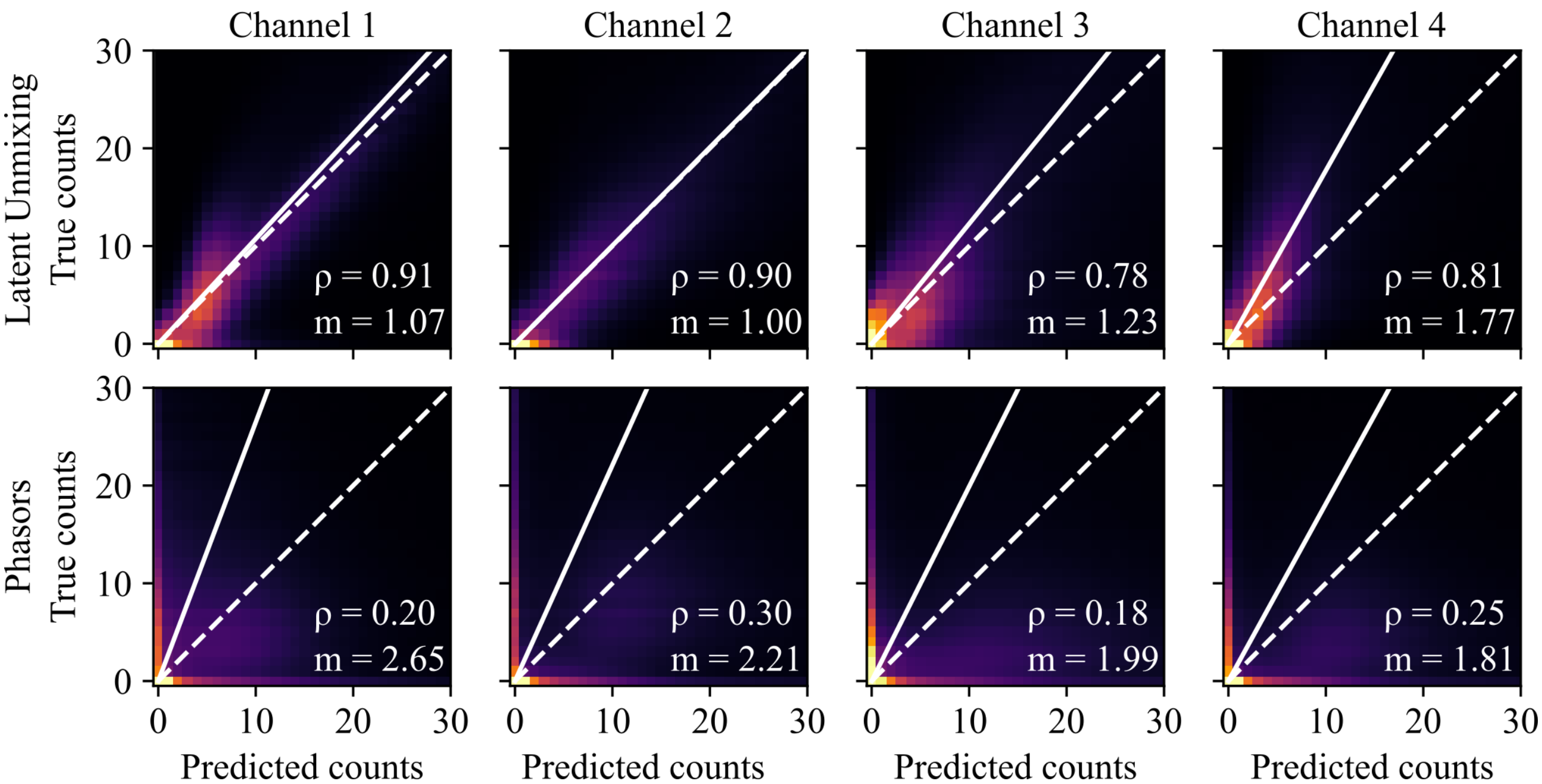

**Supplementary Figure 10 -** Correlation matrices obtained over the 12 images from the STED-FLIM test set, simulated with lifetime values of 1.0, 2.0, 3.0 and 4.0 ns, with *Latent Unmixing* (top) and Phasor analysis (bottom). ρ is the Pearson correlation coefficient, *m* is the slope of the linear regression between predicted and actual counts. Dashed lines show a perfect correlation with a slope of 1. Colormaps are square-root normalized and scaled to the second maximum of each matrix.

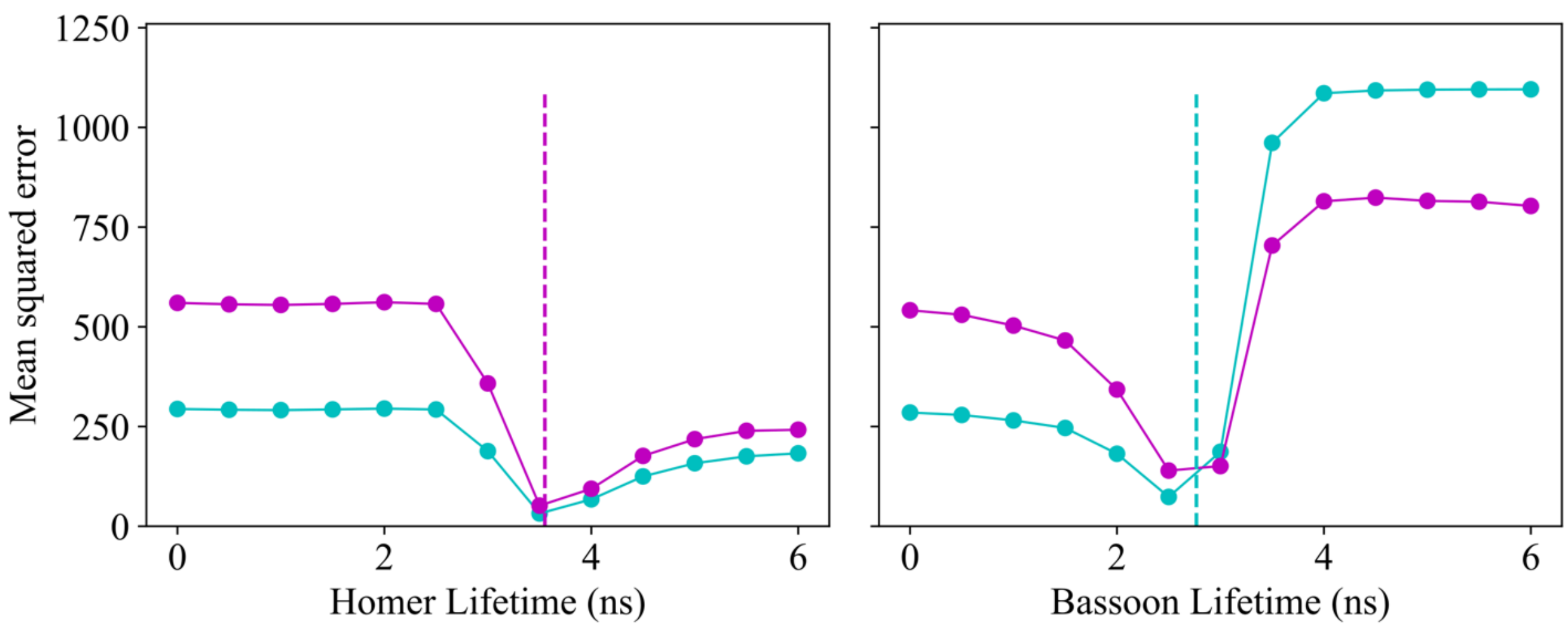

**Supplementary Figure 11 -** Mean squared error between the **unconditional** *Latent Unmixing* prediction and the ground truth for a two-channel image (Bassoon, cyan and Homer1c, magenta). The performance quickly deteriorates when the lifetime of the test data differs from the lifetime of the training data (2.77 ns for Homer1c and 3.55 ns for Bassoon, dashed lines).

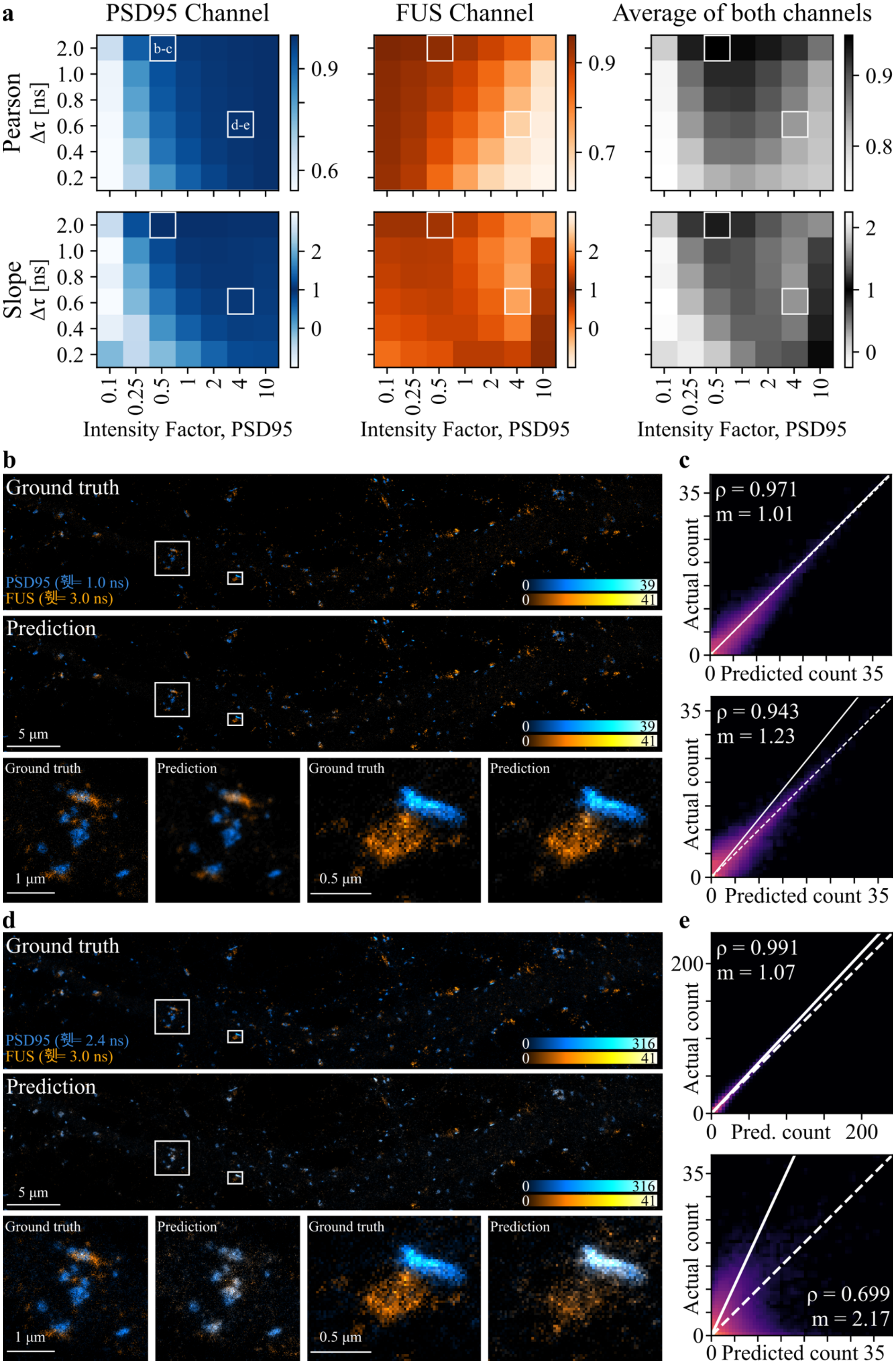

**Supplementary Figure 12 -** Optimization between the signal ratio between two channels and the lifetime proximity. **a,** Pearson correlation coefficient (top) and slope of the linear regression (bottom) between the predicted and actual photon counts for one full field-of-view test image, for different intensity factors of the PSD95 channel and lifetime differences between the two channels. For both metrics, the values closest to 1 (darker colors) are better. **b,** Ground truth and prediction for the full image and for two selected crops, for $\Delta\tau$ =2.0 ns and an intensity factor of 0.5. **c,** Correlation matrices between actual and predicted photon counts for the complete image. **d-e,** Same as **b-c** for a more difficult case defined by $\Delta\tau$ = 0.6 ns and an intensity factor of 4.

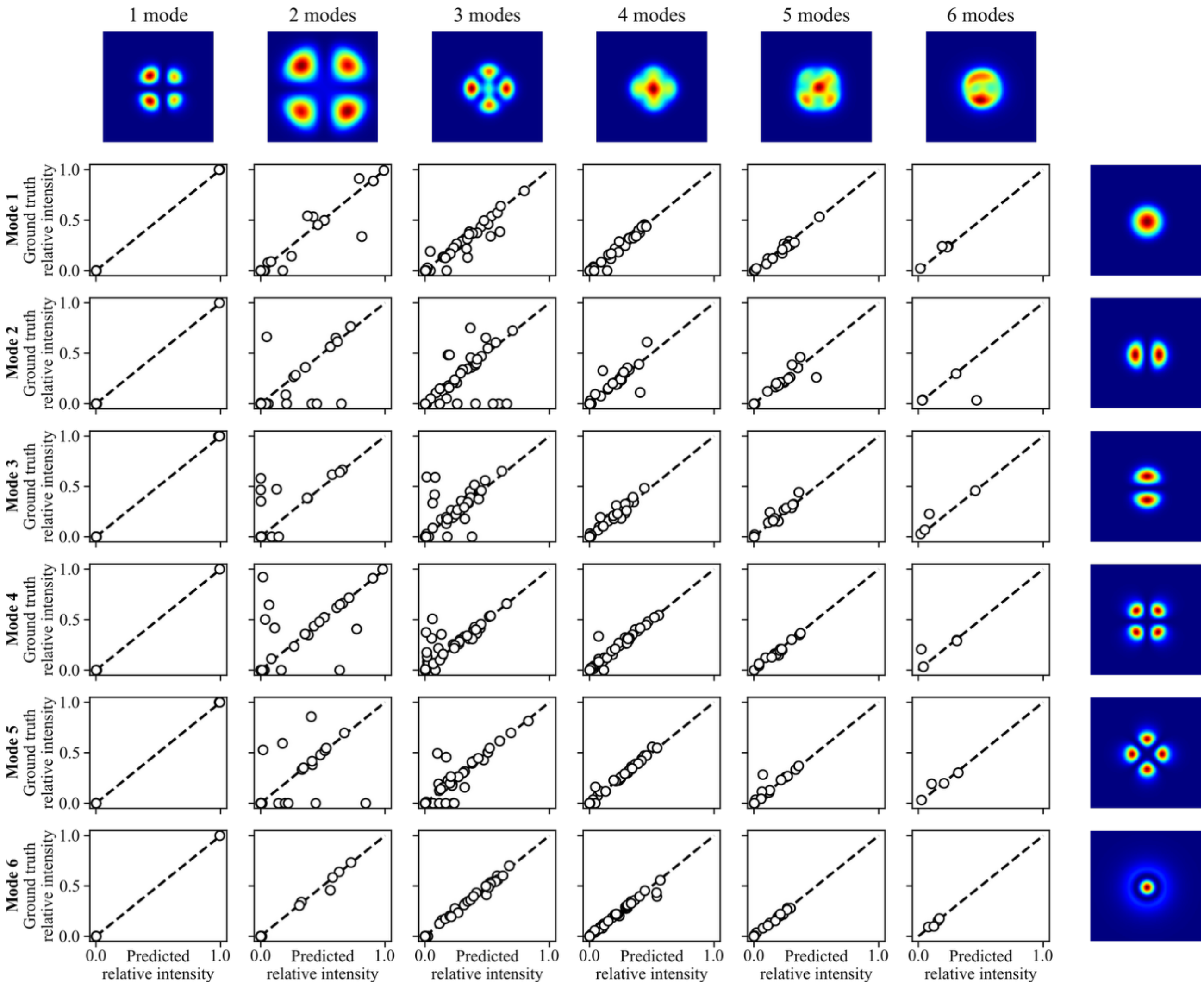

**Supplementary Figure 13 -** Correlation between the target and predicted intensity over different mixtures of propagating modes. Dashed line is the perfect correlation with a slope of 1. Columns correspond to the number of modes in the mixture (from 1 to 6) and rows correspond to a given mode (indexed from 1 to 6). The images in the first row are an example input for each number of modes, summed over the spectral dimension. The images in the last column are an example of the ground truth spatial distribution of each mode.

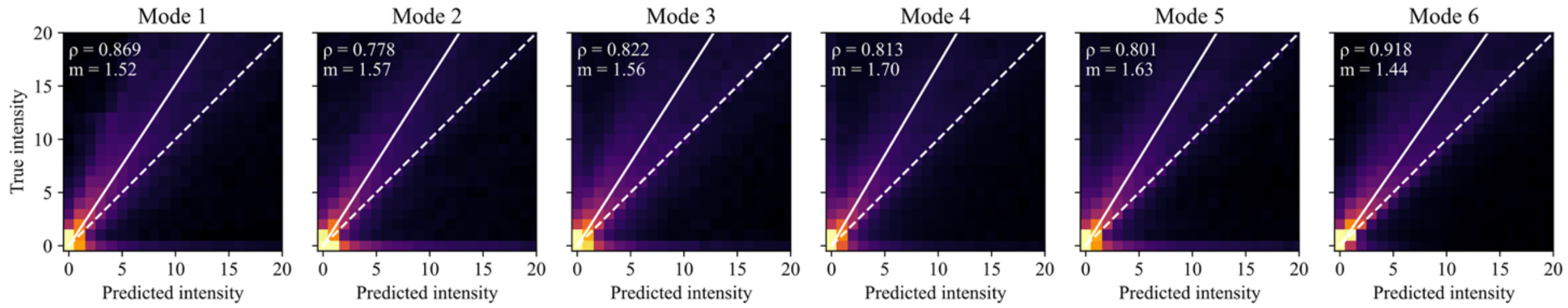

**Supplementary Figure 14 -** Correlation matrices for each mode of propagation computed over the test set (N=175 simulated measurements, 716,800 pixels). Dashed lines correspond to a perfect correlation with a slope of 1. Solid lines show the regression line of the predictions. Pearson correlation coefficients and the slope of the regression between predicted and ground truth are indicated in the top-left corner of each matrix. Colormaps are square-root normalized and scaled to the second maximum of each matrix.

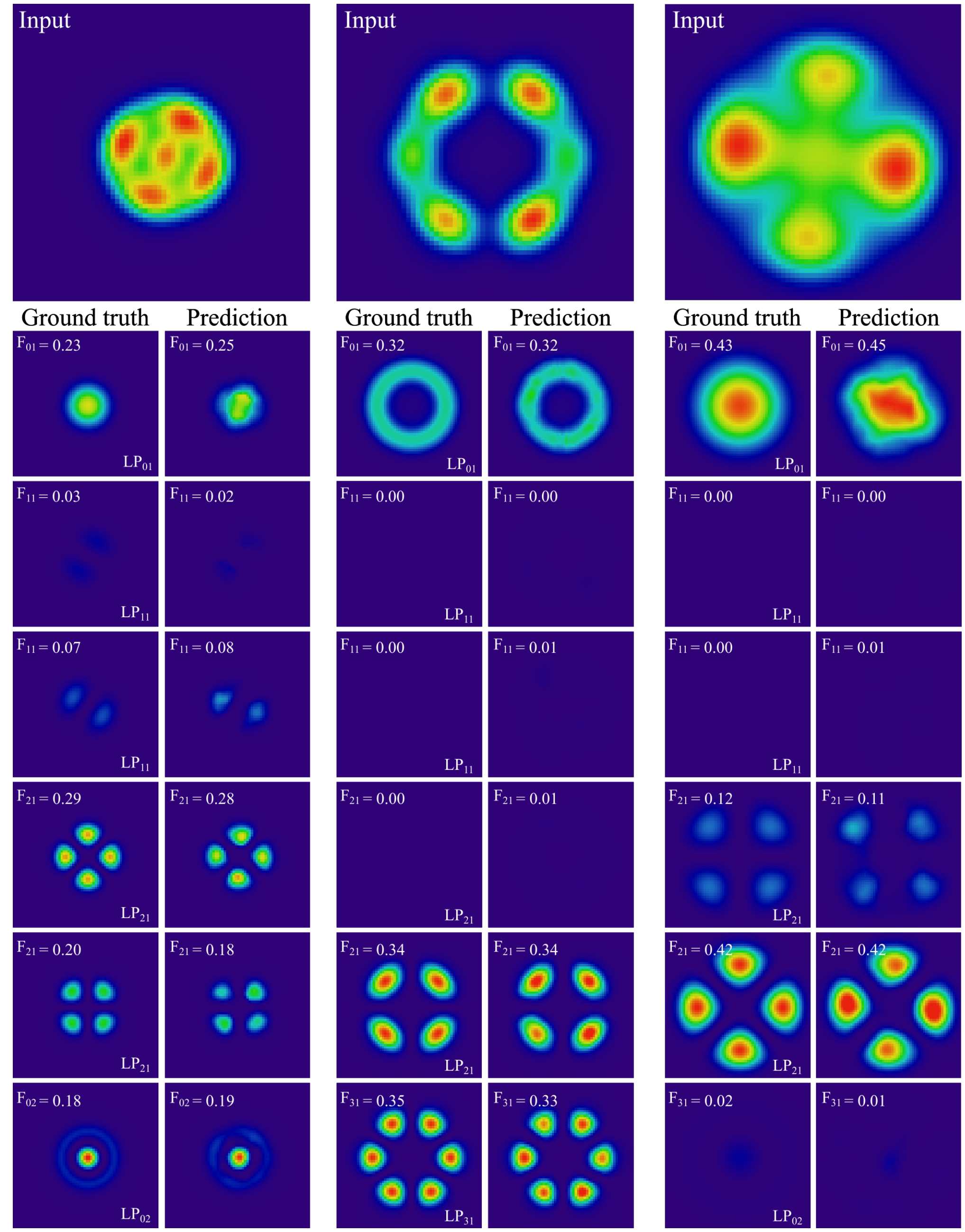

**Supplementary Figure 15 -** Ground truth and predicted spatial distribution of modes for 3 different fiber geometries from three randomly selected examples from the test set. Show as the input is the maximum projection over the spectral dimension of the actual volumetric input. The numbers at the left-upper corner of each image indicates the fraction of the total intensity in each mode, obtained by summing the pixels over the image and normalizing over the sum of the 6 images.